\documentclass[10pt]{article}
\usepackage{graphicx,amsmath,amssymb}
\usepackage{enumerate}
\usepackage{appendix}
\usepackage{setspace}

\begin{document}

\vspace{20pt}

\begin{center}
{\huge Self-interacting scalar fields at high-temperature}
\end{center}

\centerline{Alexandre Deur\,\footnote{
deurpam@jlab.org}}

\vspace{3pt}

\centerline{ \emph{University of Virginia, Charlottesville, VA 22904. USA}}

\vspace{10pt}

\begin{abstract}

We study two self-interacting scalar field theories in their high-temperature limit using path integrals on a lattice. 
We first discuss the formalism and recover known potentials to validate
the method. We then discuss how these theories can model, in the high-temperature limit, the strong interaction
and General Relativity. For the strong interaction, the model recovers 
the known phenomenology of the nearly static regime of heavy quarkonia. The model also exposes
a possible origin for the emergence of the confinement scale from the approximately conformal Lagrangian.
Aside from such possible insights, the main purpose of addressing the strong interaction here --given that
more sophisticated approaches already exist-- is mostly to further verify 
the pertinence of the model in the more complex case of General Relativity for which non-perturbative
methods are not as developed. The results have important implications on the nature of Dark Matter. 
In particular, non-perturbative effects naturally provide flat
rotation curves for disk galaxies, without need for non-baryonic matter, and explain as well other  
observations involving Dark Matter such as cluster dynamics or the dark mass of elliptical galaxies. 

%\keywords{QCD, General relativity, Dark Matter}
\end{abstract}

\section{Introduction}
The nature of confinement in a Yang-Mills theory is an important and complex question that is not settled yet. In fact, 
there is no consensus on what mechanism leads to confinement. Various candidates include center vortices, 
magnetic monopoles, Abelian Higgs mechanism, the Gribov mechanism, or the creation of an 
Abrikosov-type string \emph{via} dual Meissner effect; see~\cite{the: Greensite conf.} for a review. 
Even the phenomenology of the strong interaction's strings/flux tubes is challenged as a 
manifestation of quark confinement~\cite{Roberts:2012sv}.
What is clear is that confinement stems from field self-interactions in the non-perturbative, i.e., strong coupling, regime.

In this paper, we numerically study two self-interacting scalar field theories which may correspond, in the high-temperature limit,  to 
Quantum Chromodynamics (QCD) and General Relativity (GR). 
QCD is  the non-Abelian --and thus self-interacting -- theory of the strong interaction 
and the prototypical confining theory. Its phenomenology has
been --and continue to be-- experimentally scrutinized. Many methods have been developed to study it. They have reached 
a high degree of sophistication. Thus, our primary purpose here is not to study QCD with a simplified approach but rather
to use the strong interaction as a testing ground, once our model is checked with simpler free-field theories. 
Nevertheless, we hope that the present simple model can yield interesting insight in the confinement mechanism. 
Gravity's GR is also a self-interacting theory which ensuing non-linearities
make its resolution notoriously difficult. Studying it is not as
developed as for QCD and our model provides a relatively simple way to approach its phenomenology. The verification
of our approach in the free-field case and its relevance to QCD phenomenology, indicate that its predictions for GR
should also be pertinent. 

The paper is organized as follows. We first introduce the Lagrangians of two scalar self-interacting field theories.  
Next, the corresponding instantaneous potentials
are expressed in terms of path integrals on a lattice, and calculated with a standard numerical
Monte-Carlo technique. We then interpret the results obtained in the high-temperature regime. 
Finally, we discuss how the scalar theories may approximate the non-zero spin field theories of QCD (spin 1, i.e. vector field)
and GR (spin 2, i.e. rank-2 tensor field).  In particular, we recall how the Lagrangian
of GR can be expressed in a polynomial form which, when expressed in the static limit, is identical 
to one of the scalar Lagrangians previously introduced in the high-temperature limit. 
A first consequence of the high-temperature limit is that numerical
calculations become extremely fast and tractable. Furthermore, the high-temperature limit
takes the innate quantum nature of a path-integral formalism
to the classical limit. While this is a limitation for modeling the inherently quantum strong interaction, 
it is advantageous for approximating GR since it alleviates the notorious difficulties
associated with quantum gravity.
The strong interaction being intrinsically non-classical, there is no reason to expect that the high-temperature potential
corresponds to the known quark-quark phenomenological static potential. 
It is in fact not the case: the high-temperature (i.e. classical) potential derived with the method discussed in this article has 
approximately a Yukawa form, which is very different from the phenomenological quark-quark
linear potential. Although the classical potential cannot be compared to the phenomenological one, it is still a valuable 
result since it can be compared to results from other approaches  obtained
in the classical limit of QCD. An agreement would then validate the method used in this article. Furthermore, the
short distance quantum effects can be introduced \emph{had-oc} by allowing the theory's coupling to run. Checking whether such running
transforms the Yukawa potential into the expected linear potential would then validate further the approach and makes it relevant to 
the study of confinement. As we already made clear, the method discussed here in the context of the strong interaction
is not intended to compete with the more mature
and sophisticated approaches that are already used to study the 
strong interaction, such as Lattice QCD, the Schwinger-Dyson approach or
the Anti-de-Sitter/Conformal Field Theory (AdS/CFT) duality. Rather, the primary purpose of studying the strong interaction 
is to check the technique so that it can be applied with confidence to the more complex --but classical-- case of GR. 
A secondary benefit is that new approaches may provide fresh views on
mechanisms important to the confinement problem. As a specific example, it is interesting to see in the approach discussed here 
how a mass scale emerges out of a conformal Lagrangian. 
Such phenomenon is at the heart of recent advances in our understanding of QCD in its non-perturbative regime~\cite{Brodsky:2016yod}.
At the end of this article, after having validated our approach using the free-field and QCD cases, we discuss the predictions
of our model in the GR case. In particular, the model naturally explains many cosmological observations without need for non-baryonic 
dark matter. To show how these observations can be naturally and accurately explained by
a phenomenology well-known in the QCD context is the main goal of this article. 
These observations include the rotation curves of disk galaxies, the fact that they are approximately flat, 
The correlation between the rotation speeds and baryonic masses of disk galaxies~\cite{Tully:1977fu},
the correlation between the dark mass of elliptical galaxies and their ellipticities~\cite{Deur:2013baa}, 
the dynamics of galaxy clusters and the Bullet cluster observations~\cite{Clowe:2006eq}.

We have grouped the technical aspects
of this work in several appendices:
In Appendix~\ref{sub:Method} we recall the numerical technique used to obtain the potentials.
In Appendix~\ref{Appendix First-numerical-simulation}
we give step-by-step instructions for programming a basic calculation.
In Appendix~\ref{Appendix boundary conditions}, we discuss the question of the choice
of boundary conditions and show the independence of our results on such choices.
In Appendix~\ref{Appendix verification}, we recover several analytically known potentials to validate
the technique developed in this paper. 
In Appendix~\ref{Appendix lattice param}, we verify that the results are independent of the choice
of lattice parameters, such as mesh spacing,  decorrelation coefficient or lattice size. 
In Appendix~\ref{Appendix: Spin 0 Assumption} we recover the post-Newtonian
formalism from the scalar Lagrangian, thereby
further verifying that it correctly approximates GR in the classical limit.

This approach is simple and fast. We give in 
Appendix~\ref{Appendix First-numerical-simulation} the time necessary for a standard
computer to perform a calculation to good precision. Extending this figure to  
state-of-art lattice machines implies that a one minute calculation
would reach a relative precision of $2 \times 10^{-7}$, comparable to the intrinsic 
precision of a 32-bit machine. It should be remarked that the method and 
algorithms are not specially optimized, since the speed is adequate
for the problems discussed in this paper.

%\noindent Natural units $\hbar=c=1$ are used throughout the paper
%unless stated otherwise. The Minkowski metric signature $(+,-,-,-)$
%is used. When considering a two-source system, the line passing through
%the two sources defines the $x-$axis of the reference frame.

\section{Scalar Lagrangians}
\begin{figure}[ht!]
\begin{center}
\centerline{\includegraphics[scale=0.25, angle=0]{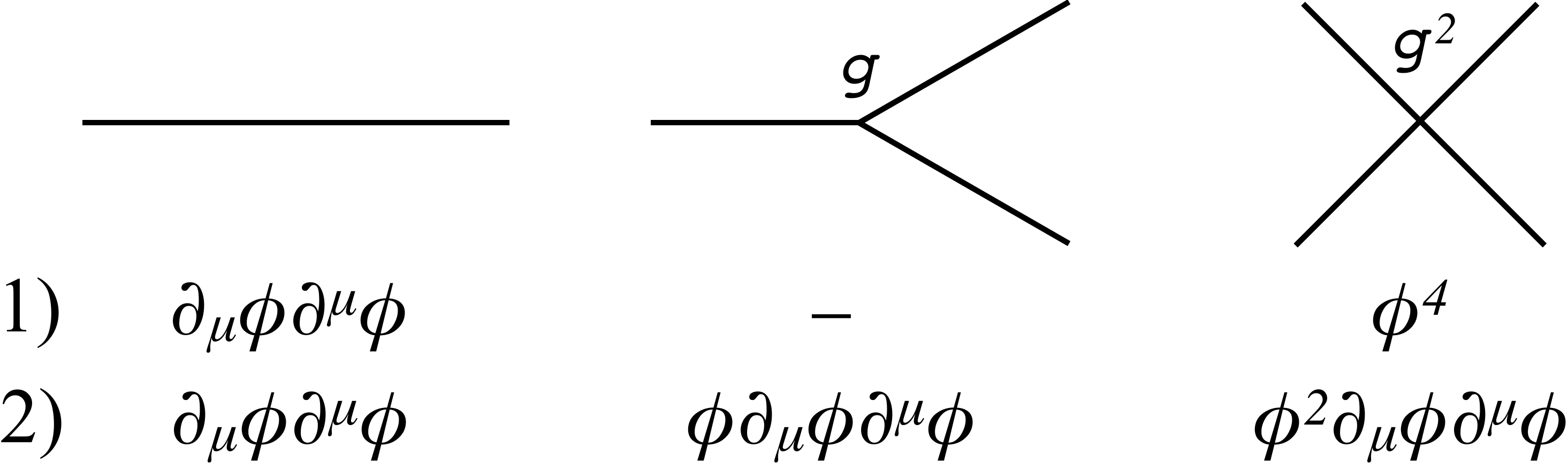}}
\end{center}
\caption{Graphs used to construct the Lagrangians. The \emph{graph on the left} corresponds to the free part of the theory yielding
 a $1/r$ potential. The \emph{two other graphs} create field self-interactions that distort this potential. 
In case (1), the forms of these terms are imposed by choosing a dimensionless coupling (no consistent cubic term can be formed).
In case (2) they are chosen for a coupling  of dimension $[$energy$]^{-2}$}
\label{fig:fg}
\end{figure}

Confinement stems from field self-interactions in large coupling or strong field regimes. 
Lagrangian densities for self-interacting theories can be constructed from
the graphs shown in Fig.~\ref{fig:fg}. We discuss here the simple case of a scalar (spin 0) field $\phi$. We consider only massless fields,
like those describing the QCD and GR forces.

The form of the Lagrangian density in case (1) is
\begin{eqnarray}
\mathcal{L}=\partial_{\mu}\phi\partial^{\mu}\phi
-g^2\phi^{4} 
\label{eq:Toy Lagrangian QCD}
,
\end{eqnarray}
with $g$ the coupling, dimensionless as for QCD.  For case (2) the Lagrangian density is
\begin{eqnarray}
\mathcal{L}=\partial_{\mu}\phi\partial^{\mu}\phi+g\phi\partial_{\mu}\phi\partial^{\mu}\phi +g^{2}\phi^{2}\partial_{\mu}\phi\partial^{\mu}\phi \label{eq:Toy Lagrangian GR} 
, 
\end{eqnarray}
with $g$ a coupling  of dimension $[$energy$]^{-2}$, as for GR. 

There is no cubic term in Eq.~(\ref{eq:Toy Lagrangian QCD}) since it would not be Lorentz invariant. 
One could insist on having such a term by adding a unit 4-vector $e^\mu$ to make the Lorentz structure of 
Eq.~(\ref{eq:Toy Lagrangian QCD}) internally consistent when a cubic term is included.  However, such  a
$g\phi^2\partial_{\mu}\phi e^{\mu}$ term would have 
no influence since it disappears from the Euler-Lagrange equation:
\begin{eqnarray}
\frac{\partial\mathcal{L}}{\partial \phi}-\partial_\mu \frac{\partial\mathcal{L}}{\partial\partial^\mu \phi}=
-2g^2 \phi^3-\partial^2 \phi,
\end{eqnarray}
as we see from the absence of a term linear in $g$. 
Only the terms stemming from the quartic component (proportional to $g^2)$ and from the quadratic one remain.

Although the scalar field Lagrangians in Eqs.~(\ref{eq:Toy Lagrangian QCD}) and~(\ref{eq:Toy Lagrangian GR}) 
stem from the same diagrams, they display different structures. There is no cubic term in Eq.~(\ref{eq:Toy Lagrangian QCD})  
and the quartic term contains no partial derivatives. Hence, it does not contribute to the propagation 
of the field and acts similarly to a mass term. In fact, the analytical solution for the equation of motion is known, 
with the field obeying a massive dispersion relation~\cite{Frasca:2009bc}, whose resulting potential can be 
well fit by a Yukawa potential, as we show in Sect.~\ref{sec:Interpretation QCD}. In Eq.~(\ref{eq:Toy Lagrangian GR}),
 the cubic term is preserved. All terms contain partial derivatives and thus contribute to the field propagation.

\section{Computation of the potential between two static sources \label{sub:numerical calc.}}
The potential at a point $x_{2}$ from a point-like
source at $x_{1}$ can be obtained, in the high-temperature limit, by the two-point Green function $G_{2p}(x_{1}-x_{2})$.
In the Feynman path-integral formalism $G_{2p}$ is
\begin{equation}
G_{2p}(x_{1}-x_{2})=\frac{1}{Z}\intop\mathrm{D}\varphi\,\varphi(x_{1})\varphi(x_{2})\mathrm{e}^{-\mathrm{i}\, S_{\mathrm{s}}},\label{eq:2pt green}\end{equation}
where $S_{\mathrm{s}}\equiv{\int{\mathrm{d}^{4}x\,\mathcal{L}}}$
is the action, $\intop{\mathrm{D}\varphi}$ sums over all possible
field configurations, and $Z\equiv\intop{\mathrm{D}\varphi\,\mathrm{e}^{-\mathrm{i}\, S_{\mathrm{s}}}}$. On an 
Euclidean spacetime lattice simulation, the 
temperature corresponds to inverse of the lattice spacing in the time direction. Hence, in the high-temperature limit, 
the sum $\intop{\mathrm{D}\varphi}$ is reduced to configurations in position space. The suppression
of time allows us to identify the potential to $G_{2p}$.

Such expression of the potential is unconventional and restrictive since it applies only to stationary systems. 
However, its simplicity allows for fast calculations, which in turn may allow the implementation of forces 
too complex, e.g. gravity with tensor fields, to be efficiently implemented on the lattice in the standard 
way. Since such a way of calculating potentials is unusual, we will specifically verify its validity  for  instantaneous potentials, 
the only cases studied here. This will be demonstrated by correctly recovering potentials in known cases of free-field theories 
(theories without self-interacting fields). Although adding terms to the Lagrangian does not affect the above definition of the potential,  
one might speculate that self-interacting terms may void its validity. 
However, we will also recover the known expected potential for the self-intercating $\phi^4$ theory
as well as the  strong interaction phenomenological static potential once short distance quantum effects are accounted for.
Based on this we can assume with some confidence that this method of calculating potentials remains valid for 
the self-interacting field case.

\subsection{Quantum versus classical calculations \label{sub:Quantum-vs-classical}}
(In this section we keep $\hbar$ in expressions to keep track of
quantum effects.)
\noindent In principle, the path-integral formalism provides intrinsically
quantum calculations. However, the results obtained here will be classical since we 
employ a lattice covering position space only (high-temperature limit)~\cite{Buchmuller:1997nw}.
Briefly, the system being time-independent, we have $S_{\mathrm{s}}\equiv{\int{\mathrm{d}^{4}x\,\mathcal{L}}=\tau S}$,
with $\tau=\intop_{t_{0}}^{\infty}{\mathrm{d}t\rightarrow\infty}$
and $S\equiv{\int{\mathrm{d}^{3}x\,\mathcal{L}}}$.
The exponential term in Eq.~(\ref{eq:2pt green}) becomes 
$\mathrm{e}^{\mathrm{-i}S_{\mathrm{s}}/\hbar}=\mathrm{e}^{\mathrm{-i}\tau S/\hbar}$.
Compared to the usual $\mathrm{e}^{\mathrm{-i}S/\hbar}$ exponential in path-integral quantum field
theory, quantum effects are suppressed
as $\hbar/\tau$. Since $\tau\rightarrow\infty$, this is 
equivalent to the classical $\hbar  \rightarrow 0$ limit. 
%Thus, the results discussed here will be classical. 

We remark that our definition of the potential allows one to compute 
only instantaneous potentials, which are suited for stationary systems and 
the purpose of the present article. Consequently, computing potentials from the time-propagation of the system, as for example 
in the Feynman-Kac Formula~\cite{Feynman:1948ur,Kac} or using Wilson loops~\cite{Wilson:1974sk}, is not applicable.

In the rest of the paper, we will not write anymore the factor $\tau$ since, as a constant rescaling factor,  it is irrelevant to 
classical calculations (a rescaled extremal action remains extremal or, equivalently, rescaling $S$ amounts to rescaling 
$\mathcal{L}$, whose rescaling factor cancels out in the Euler-Lagrange Equation). 
However, $\tau$ must be considered when e.g. discussing the dimensions of the fields and couplings.
In other words, the action $S$ has a dimension $[\hbar \times $energy$]$ rather than $[ \hbar ]$.

\subsection{Goal, formalism and verifications}
The goal is to compute the potential between two approximately
static ($v\ll c$) sources in the non-perturbative regime. 
The lattice technique, based on a
standard Monte-Carlo method exploiting the Metropolis algorithm,  is described in Appendix~\ref{sub:Method}.
A practical example is given in Appendix~\ref{Appendix First-numerical-simulation}.
In Appendix~\ref{Appendix boundary conditions}, we discuss the delicate point of the choice of
boundary conditions and show that the results are independent of such choices.  In Appendix~\ref{Appendix verification}, the method is verified by recovering analytically known 
free-field potentials in two or three spatial dimensions,  as well as the $g^2 \phi^4$ theory potential in three dimensions. In
Appendix~\ref{Appendix lattice param}, we verify the independence of the results on the choice
of lattice parameters.

\section{Results}
\subsection{$g^2 \phi^4$ theory \label{sec:Interpretation QCD}}
\begin{figure}[ht!]
\centering
 \includegraphics[width=0.6\textwidth]{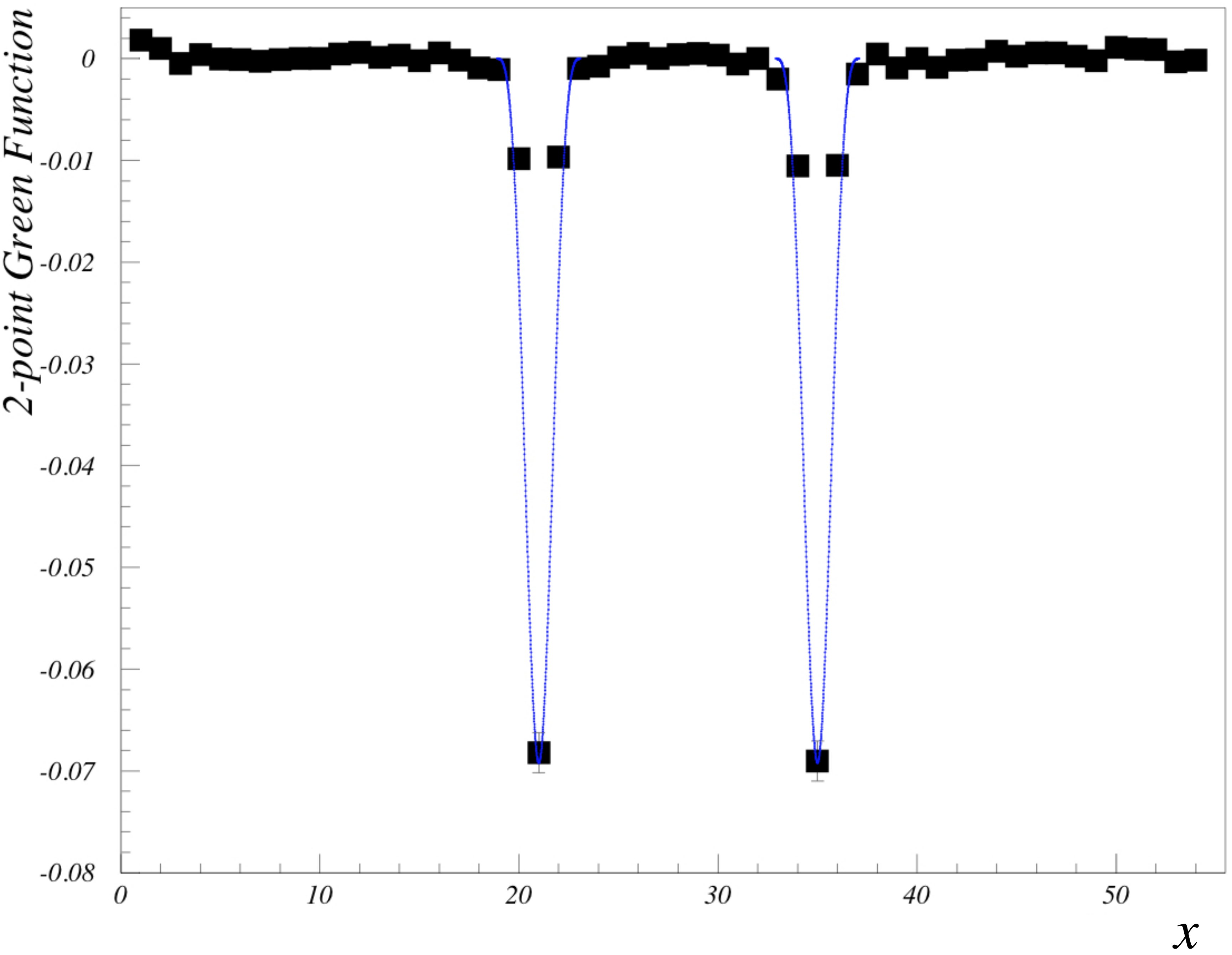}
\caption{
\label{fig:QCD_sn_check}Comparison between the  potential computed numerically
using Eq.~(\ref{eq:Toy Lagrangian QCD}) (\emph{squares}) and the analytical 
expectation $\mbox{sn}(x,-1)\star \mbox{sn}(x,-1)$ (\emph{line}).
The simulation used a coupling $g^2 =3$, a lattice size $N=55$, $N_{\mathrm{s}}=5\times 10^{4}$ configurations,
a decorrelation coefficient $N_{\mathrm{cor}}=15$ and random field boundary conditions
\protect \\
}
\end{figure}

The $g^2 \phi^4$ theory can be solved analytically~\cite{Frasca:2009bc}. The analytical solution of the equation 
of motion is known,  with the field obeying a massive dispersion relation. Indeed,
the resulting potentials are well fit by a Yukawa potential over the whole range of  coupling values we used
($0\leq g^2 \leq 4$); see Fig.~\ref{Flo:Boundary conditions, qcd case} for an example 
of potential between two sources located at $x_1=21$ and $x_2=35$, and for $g^2 = 0.5$. 
Figure~\ref{Flo:Boundary conditions, qcd case} also demonstrates the independence of the result with the choice of boundary conditions.
The expected function describing $\phi$ is the Jacobi elliptic function $\mbox{sn}(x,-1)$~\cite{Frasca:2009bc}. Its
autocorrelation $\mbox{sn} \star \mbox{sn}$, which provides 
the potential $G_{2p}$, fits well our numerical results; see Fig.~\ref{fig:QCD_sn_check}, 
but only for coupling values $g^2 > 1$.  

\begin{figure}[ht!]
\centering
 \includegraphics[width=0.6\textwidth]{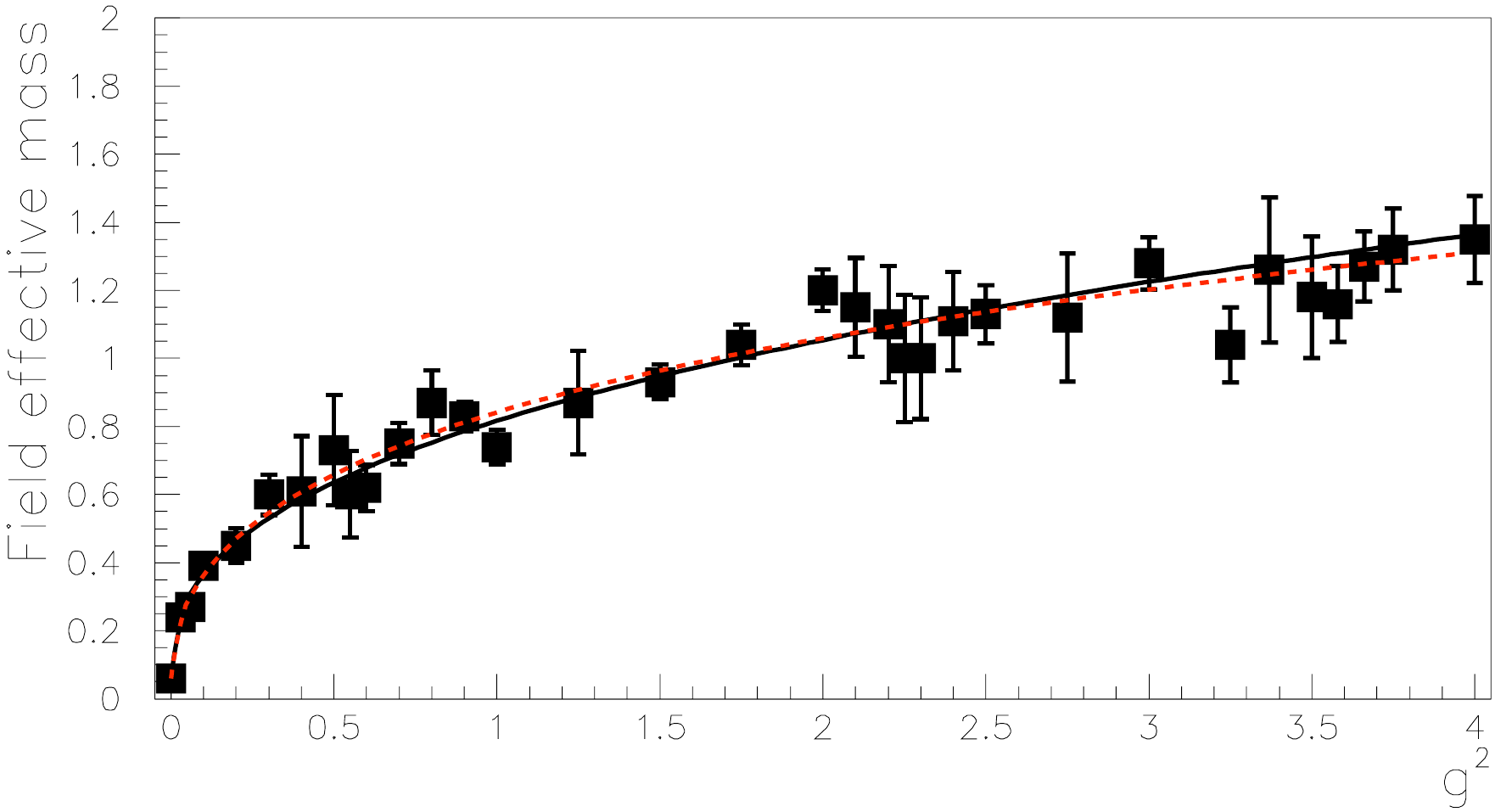}
\caption{
\label{fig:mass vs alpha_s}The  $g^2$-dependence of the effective mass generated by the quartic term $g^2 \phi^4$.
The \emph{continuous line} is a polynomial fit to the data and the \emph{dashed line}
 is a logarithmic fit. The simulation used a lattice size $N=55$, $N_{\mathrm{s}}=5\times 10^{4}$ configurations,
a decorrelation coefficient $N_{\mathrm{cor}}=15$ and random field boundary conditions
\protect \\
}
\end{figure}

The effect of the quartic term $g^2 \phi^4$ can be embodied using a field effective mass $m_{\mbox{\scriptsize{eff}}}$. It can be obtained
by fitting $G_{2p}$ with a Yukawa potential $e^{-m_{\mbox{\scriptsize{eff}}}~x}/x$. Such effective mass
is shown as a function of $g^2$ in Fig.~\ref{fig:mass vs alpha_s}. 
A polynomial fit to these data  yields $m_{\mbox{\scriptsize{eff}}} = (0.76\pm0.02)( g^2)^{0.39\pm0.02}+0.05\pm0.02$, 
and a logarithmic fit yields
$m_{\mbox{\scriptsize{eff}}} = (1.13\pm0.03)\mbox{ln}\big((g^2)^{0.51\pm0.02}+1\big)+0.06\pm0.02$ with a marginally better $\chi^2$ (10\% smaller). 
As already mentioned, the effectively massive behavior of the $g^2 \phi^4$ theory can be 
understood from the lack of partial derivatives in the quartic term. It thus does not contribute 
to the field propagation and is akin to a mass term. Evidently, this field effective mass depends on
the value of the coupling $g^2$ (Fig.~\ref{fig:mass vs alpha_s}). 
\begin{figure}[ht!]
\centering
 \includegraphics[width=0.6\textwidth]{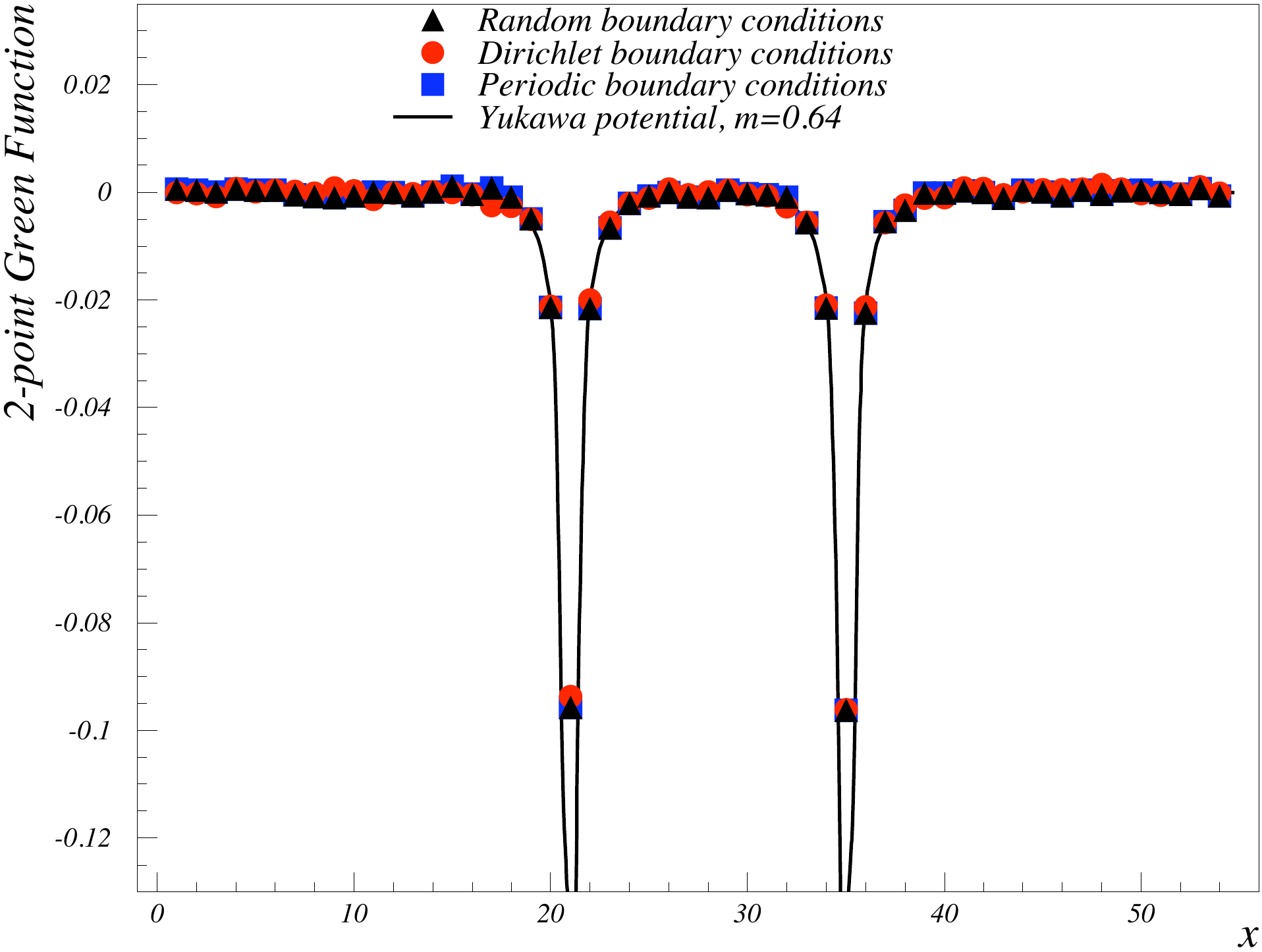}
\caption{
\label{Flo:Boundary conditions, qcd case} Potential
$G_{2p}$ for the $g^2 \phi^4$ theory, plotted as a function of distance. The simulations used
a lattice size $N=55$, a coupling $g^2=0.5$, a decorrelation
coefficient (see Sect.~\ref{sub:First-simulation}) $N_{\mathrm{cor}}=20$
and $N_{\mathrm{s}}=5\times10^{4}$ configurations. The results are independent of
the choice of boundary conditions. They are
well fit by two Yukawa potentials centered 
on each two sources at $d=\pm 7$ from the lattice center and with a mass 
parameter $m_{\mbox{\scriptsize{eff}}}=0.64$ (\emph{continuous line}) \protect \\
}
\end{figure}
\subsection{($g \phi \partial_\mu \phi \partial^\mu \phi + g^2 \phi^2 \partial_\mu \phi \partial^\mu \phi$) theory}
\begin{figure}[ht!] 
\centering
 \includegraphics[width=0.6\textwidth]{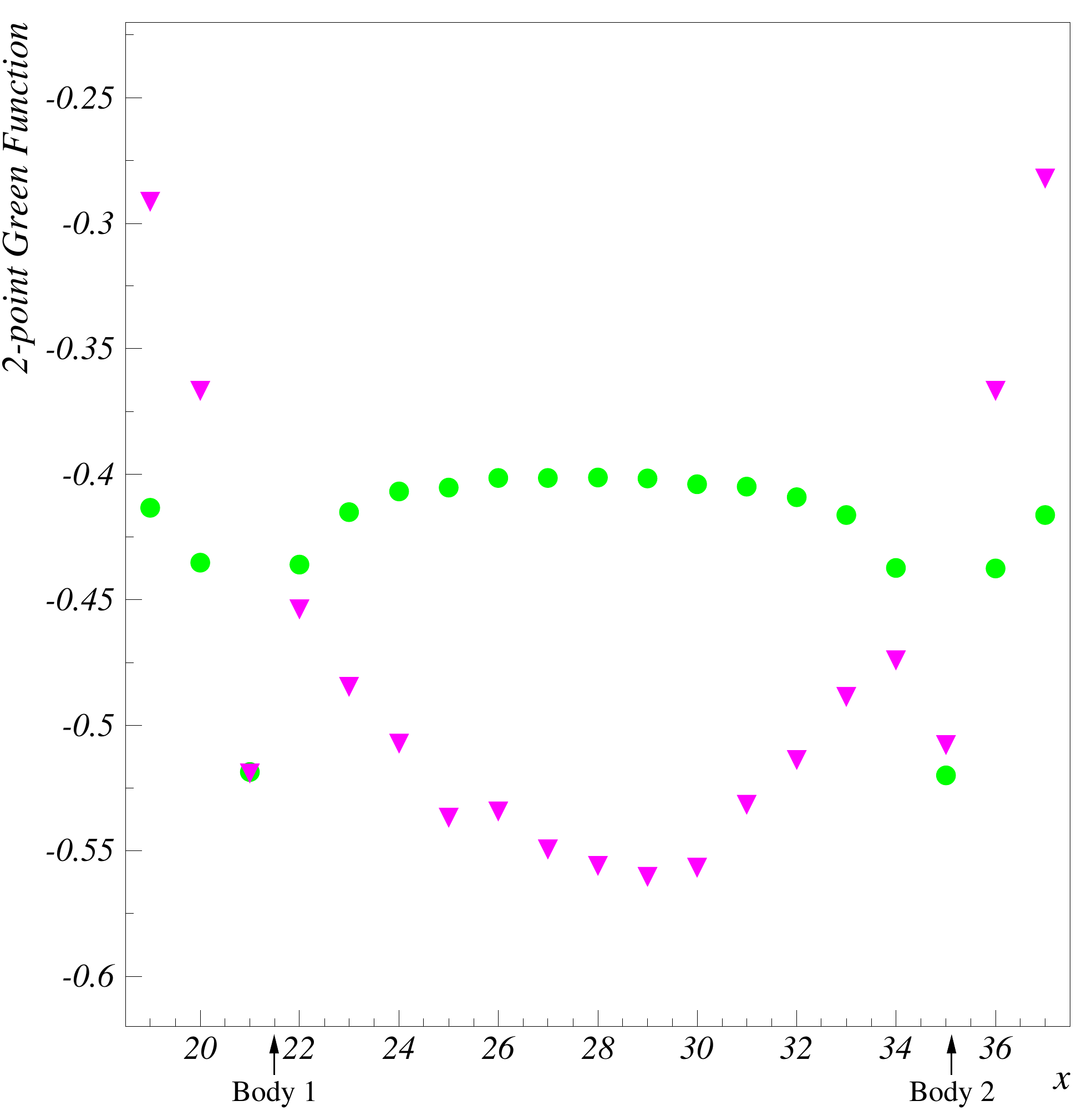}
\caption{\label{Flo:size effect} The potential for the 
($g \phi \partial_\mu \phi \partial^\mu \phi + g^2 \phi^2 \partial_\mu \phi \partial^\mu \phi$) theory  (\emph{triangles}). 
The \emph{circles} are the free-field case ($g=0$). The two sources are located on the $x-$axis
at $d=\pm7$ from the lattice center $x_{\mathrm{center}}=28$, $y=0$
and $z=0$. The coupling is $g=7.5\times10^{-3}$, the lattice size is $N=85$, the decorrelation parameter is $N_{\mathrm{cor}}=20$ and 
$N_{\mathrm{s}}=3.5\times10^{4}$ paths were used.
To conveniently compare the results, a constant offset is added to
the free-field case so that the potentials at the position of the first source, $x=21$, match \protect \\
}
\end{figure}

\begin{figure}[ht!] 
\centering
 \includegraphics[width=0.6\textwidth]{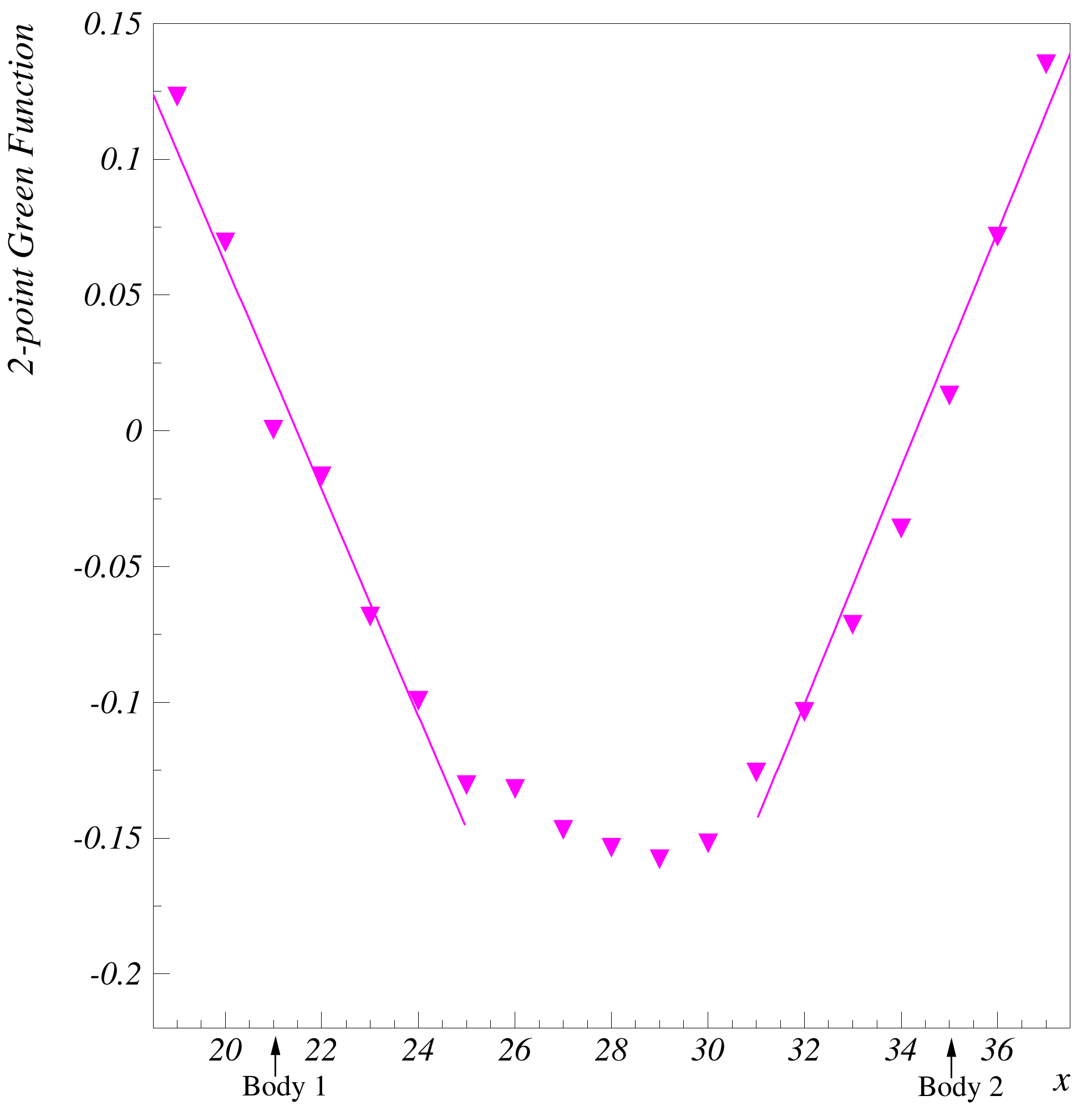}
\caption{
\label{Flo:size effect Newton subtracted}Same as Fig.~\ref{Flo:size effect}
but with the $1/x$ --free-field--  contribution subtracted. The \emph{lines}
illustrate the nearly linear trend of the potential\protect \\
}
\end{figure}

The ($g \phi \partial_\mu \phi \partial^\mu \phi + g^2 \phi^2 \partial_\mu \phi \partial^\mu \phi$) theory 
in its non-perturbative regime yields the static potential shown in Fig.~\ref{Flo:size effect}. It varies approximately linearly with distance (except
at mid-distance between the two sources where the potential flattens out, as expected for a solely attractive force).
Also shown is the $g=0$ case which yields the expected free massless field potential in $1/x$, with $x$ the distance to one of the source. 
In Fig.~\ref{Flo:size effect Newton subtracted}, results are shown with the $1/x$ contribution subtracted in order to isolate the linear part stemming 
from the  ($g \phi \partial_\mu \phi \partial^\mu \phi + g^2 \phi^2 \partial_\mu \phi \partial^\mu \phi$) term. 

The linear $x$-dependence of the potential can be interpreted in a similar
manner as done in the QCD static case; see e.g.~\cite{the: Greensite conf.}. In
QCD, the linear potential is often pictured as originating from the collapse of the force's field lines into
a flux tube linking the two sources. This collapse is caused by strong 
non-perturbative field self-interactions: the  gluons strongly attract
each other toward the region of their highest density i.e. the
segment joining the sources. If the sources are static and since the
field has collapsed into a segment, the two dimensions transverse to
the segment become irrelevant. The strong non-perturbative
effects have constrained the field flux from three spatial degrees
of freedom to one. The resulting force is constant, as
easily calculated or visualized with Faraday's flux picture.

A similar phenomenon seems to occur with the ($g \phi \partial_\mu \phi \partial^\mu \phi + g^2 \phi^2 \partial_\mu \phi \partial^\mu \phi$)
theory. The strong field self-interaction effectively
reduces the three-dimensional system to a one-dimensional system,
yielding a linear potential. The linear potential can also be understood from  field correlations. The potential
is given by the two-point Green function, i.e. the autocorrelation
function of the field; see Eq.~(\ref{eq:2pt green}). It is a general property
that the autocorrelation of the sum of two uncorrelated functions
is the sum of the autocorrelations of each function separately: if
$f=g+h$, then $f\star f=g\star g+h\star h$, i.e. two uncorrelated
 \emph{fields} $g$ and $h$ (i.e. $g$ and $h$ do not
influence each others) from two sources imply the superposition principle.
The total \emph{potential} $f\star f$ is the sum of the potentials
$g\star g$ and $h\star h$. If the fields mutually influence each
others, correlations appear. The superposition
principle is thus broken.
When fields are interacting strongly enough, they
become strictly correlated. The total field (observed sufficiently
away from the sources so that the three-dimensional symmetry around
a point-like source is broken) is then constant on a segment linking
the two sources. It must quickly go to zero outside the segment due to
the boundary conditions at infinity. The field is then approximately
a one-dimensional rectangular function. The autocorrelation of a rectangular
function being a triangle function, the potential between the two
sources has a linear $x$-dependence, as seen in Fig.~\ref{Flo:size effect Newton subtracted}.
In all, the linear potential arises due to the long field autocorrelation length --stemming from
the partial derivatives present in each term of the Lagrangian-- that couples the field at different locations in space. 
This explanation is consistent with the naive dimensional argument inspired from QCD. 

With this interpretation, we can immediately generalize the present two-source linear potential
result to a uniform and homogeneous disk of sources. The three-dimensional
system is then reduced to two dimensions. This yields a logarithmic potential (see Fig~\ref{fig: 2D abelian case}).
For a uniform and homogeneous spherical distribution of sources, the system stays three-dimensional and retains a static $1/x$ potential. 

\section{Connections to QCD and GR}
In this section, we explore the possible connections of the Lagrangians in 
Eqs.~(\ref{eq:Toy Lagrangian QCD}) and~(\ref{eq:Toy Lagrangian GR})
with QCD and GR, respectively. We first discuss the benefits that such connections would provide. 

Numerical studies of QCD using lattice gauge theory are sophisticated and well checked. They have provided invaluable information
on this theory. As stated in the introduction, the present work is not meant to compete with these more exact and more mature approaches. 
The primary goal of making the connection to QCD is that, if the QCD phenomenology
is recovered, it would support the validity of the method beyond the checks already discussed. 
The method can then be used with confidence to address more complicated cases, 
e.g. GR whose tensorial field makes it currently impractical to put on the lattice.
This primary goal being clarified, there are still advantages of such approach within the sole context of the strong interaction: 
(1) a simple approach showing the onset of confinement may isolate its essential aspects and make them easier to study, including analytically. 
Furthermore,
(2) if the confinement mechanism is manifest in a simplified approach, it may provide guidances for an analytical resolution of the problem.
It could (3) also provide an alternate approach to a complex problem. In fact, lattice QCD has greatly benefited from other non-perturbative approaches
that can be fully solved only numerically, e.g. the Schwinger-Dyson equations approach.
(4) It makes calculations fast compared to typical lattice running time --a few hours of running on a standard personal computer are sufficient.
(5) It provides a pedagogical model of confinement.

Efforts to solve GR in its non-perturbative regime by numerical methods also constitute an active field of study~\cite{Lehner:2014asa}.
Its object is strongly-interacting systems for which
gravitational backreaction cannot be ignored. As in QCD, the equations governing
these systems are non-linear and can be non-perturbative. Except in rare
cases, they can only be solved numerically. A non-perturbative component to the potential would not be 
identified using the usual perturbative (post-Newtonian) approximation, and a non-perturbative linear contribution to the
potential is in general compatible with the field equations of bound systems~\cite{Hoyer:2009ep,Dietrich:2012un}. In fact, it is emphasized
in Refs.~\cite{Hoyer:2009ep,Dietrich:2012un} that bound states can be understood 
semi-classicaly as divergences of the perturbative expansion arising
from ladder-type Feynman diagrams.
Consequently, even more than for QCD, possessing alternate methods to approach 
the complex problem of bound systems in GR  is beneficial. 

In the rest of this section, we discuss the possibility of such method, based on the fact that
the Lagrangians in Eqs.~(\ref{eq:Toy Lagrangian QCD}) and~(\ref{eq:Toy Lagrangian GR}) may be related to the 
QCD and GR Lagrangians. Such relation has already been put forward  in the case of QCD~\cite{Frasca:2007uz,Frasca:2009yp} where it was 
asserted that Eq.~(\ref{eq:Toy Lagrangian QCD}) is equivalent to QCD in either the classical limit or 
the low energy limit.  We will show that similarly,  Eq.~(\ref{eq:Toy Lagrangian GR}) may  describe GR in the 
classical limit. We first start by recalling the Lagrangians of QCD and GR.  

\subsection{QCD Lagrangian}
The QCD field Lagrangian is
\begin{eqnarray}
\label{eq:QCD Lagrangian}
\mathcal{L}^{field}_{QCD}=-\frac{1}{4}F_{\mu \nu}^a F^{\mu \nu}_a= 
\frac{1}{4}\big(\partial_{\nu}A_\mu^a - \partial_{\mu}A^a_\nu  \big) \big(\partial^{\mu}A^{\nu a} - \partial^{\nu}A^{\mu a}  \big)  \\
+\sqrt{\pi \alpha_s}f^{abc}\big(\partial_{\nu}A_\mu^a - \partial_{\mu}A_\nu^a\big)A^{\mu b}A^{\nu c} 
-\pi \alpha_s f^{abe}f^{cde}A^a_\mu A^b_\nu A^{\mu c} A^{\nu d}, \nonumber 
\end{eqnarray}
with $\alpha_s$ the QCD coupling constant, $A_\mu^a$ the gluon field  and with the SU(3) index $a=1,\ldots, 8$. 
We do not include Fadeev-Popov ghosts in the Lagrangian. Their necessity  depends on the choice of gauge, e.g. there are
no ghosts in the axial gauge $e^\mu A^a_\mu$ or in the  light-cone gauge $A^+=0$. Furthermore, Fadeev-Popov ghosts are needed in
QCD essentially to force the gluon propagator to remain transverse when loop diagrams are included and
since we are discussing here limits using spinless fields, there is no such need.  
Fermions are also not included in Eq.~(\ref{eq:QCD Lagrangian}) since it is sufficient 
to work in the pure field sector for a first study of confinement in the static case.

\subsection{General relativity}
Einstein's field equations of GR are generated by the Einstein-Hilbert
Lagrangian density: 
\begin{equation}
\mathcal{L}_{\mathrm{GR}}=\frac{1}{16\pi G}\sqrt{\mathrm{det}(g_{\mu\nu})}\, g_{\mu\nu}R^{\mu\nu},\label{eq:Einstein-Hilbert Lagrangian}
\end{equation}
where $G$ is the Newton constant, $g_{\mu\nu}$ the metric and $R_{\mu\nu}$ the Ricci
tensor. The gravity field $\varphi_{\mu\nu}$ is defined as the difference
between $g_{\mu\nu}$ and a constant reference metric $\eta_{\mu\nu}$:
$\varphi_{\mu\nu}\equiv\left(g_{\mu\nu}-\eta_{\mu\nu}\right)/\sqrt{16\pi G}$.
Here, $\eta_{\mu\nu}$ will be the flat metric. Expanding $\mathcal{L}_{\mathrm{GR}}$
in terms of $\varphi_{\mu\nu}$ yields~\cite{Salam/Zee}:
\begin{eqnarray}
\mathcal{L}_{\mathrm{GR}}={\left[\partial\varphi\partial\varphi\right]+}\sqrt{16\pi G}\left[\varphi\partial\varphi\partial\varphi\right]+16\pi G\left[\varphi^{2}\partial\varphi\partial\varphi\right] + \label{eq:Polynomial Einstein-Hilber Lagrangian} \nonumber  \\  
\cdots  -\sqrt{16\pi G}\,\varphi_{\mu\nu}T^{\mu\nu}-\frac{16\pi G}{2}\varphi_{\mu\nu}\varphi_{\sigma\lambda}T^{\mu\nu}\eta^{\sigma\lambda}+\cdots,\end{eqnarray}
where $T^{\mu\nu}$ is the energy-momentum tensor and $\left[\varphi^{n}\partial\varphi\partial\varphi\right]$
is a shorthand notation for a sum over the possible Lorentz invariant
terms of the form $\varphi^{n}\partial\varphi\partial\varphi$. For
example, $\left[\partial\varphi\partial\varphi\right]$ is explicitly
given by the Fierz-Pauli Lagrangian~\cite{Fierz-Pauli}, the
first order approximation of GR that leads to Newton's gravity:
\begin{eqnarray}
\left[\partial\varphi\partial\varphi\right]=\frac{1}{2}\partial^{\lambda}\varphi_{\mu\nu}\partial_{\lambda}\varphi^{\mu\nu}-\frac{1}{2}\partial^{\lambda}\varphi_{\mu}^{\mu}\partial_{\lambda}\varphi_{\nu}^{\nu}-
\partial^{\lambda}\varphi_{\lambda\nu}\partial_{\mu}\varphi^{\mu\nu}+\partial^{\nu}\varphi_{\lambda}^{\lambda}\partial^{\mu}\varphi_{\mu\nu}.
 \label{eq:Fierz-Pauli Lagrangian} 
\end{eqnarray}

Equation~(\ref{eq:Polynomial Einstein-Hilber Lagrangian})
is expanded in term of the dimensionless quantity $\sqrt{16\pi G}\varphi$. 
It is justified to truncate
 Eq.~(\ref{eq:Polynomial Einstein-Hilber Lagrangian}),
even for strong gravity fields, since in general $\sqrt{16\pi G}\ll1/\mu$
with $\mu$ the typical mass scale or inverse distance scale of the
system considered. A pertinent reference metric $\eta_{\mu\nu}$ is
chosen according to the system being considered. It can be e.g. 
the Minkowski metric for systems inducing weak curvatures, or the
Schwarzschild metric for a system involving a black hole. Hence,
Eq.~(\ref{eq:Polynomial Einstein-Hilber Lagrangian}) is applicable to both the weak-field
approximations of GR and non-perturbative gravity.

\subsection{Lagrangians in the static case \label{sec: Static Lagrangian}}

Equation~(\ref{eq:Polynomial Einstein-Hilber Lagrangian})
can be simplified. % in the static limit. 
Consider first the $[\partial\varphi\partial\varphi]$
term given by Eq.~(\ref{eq:Fierz-Pauli Lagrangian}). The Euler-Lagrange
equation obtained by varying $\varphi_{\mu\nu}$ in 
$\int{\mathrm{d}^{4}x\big(\big[\partial\varphi\partial\varphi\big]-}$
$\sqrt{16\pi G}\,\varphi_{\mu\nu}T^{\mu\nu}\big)$
leads to 
$\partial^{2}\varphi^{\mu\nu}=-16\pi G\big(T^{\mu\nu}-\frac{1}{2}\eta^{\mu\nu}\mbox{Tr}~T\big)$.
Since $T^{00}$ dominates over the other components of $T^{\mu\nu}$
in the static limit, $\partial^{2}\varphi^{\mu\mu}\gg\partial^{2}\varphi^{\mu\nu}$
for $\mu\neq\nu$ and $\partial^{2}\varphi^{00}\simeq\partial^{2}\varphi^{ii}$.
Thus, $\varphi^{00}=\varphi^{ii}+a_{i}x+b_{i}$. Since $\varphi^{\mu\nu}\rightarrow0$
for $x\rightarrow\infty$, $a_{i}=0$, $b_{i}=0$ and $\varphi^{00}=\varphi^{ii}$.
Replacing the $\varphi^{ii}$ terms in Eq.~(\ref{eq:Fierz-Pauli Lagrangian})
by $\varphi^{00}$ and applying the harmonic gauge condition $\partial^{\mu}\varphi_{\mu\nu}=\frac{1}{2}\partial_{\nu}\varphi_{\kappa}^{\kappa}$,
leads to $\left[\partial\varphi\partial\varphi\right]=\partial_{\lambda}\varphi^{00}\partial^{\lambda}\varphi_{00}$.
Consider now the next order term, $[\varphi\partial\varphi\partial\varphi]$.
The $\varphi^{\mu\nu}$ $(\mu\neq\nu)$ terms that were neglected
in the lower order operations leading to 
$\left[\partial\varphi\partial\varphi\right]=\partial_{\lambda}\varphi^{00}\partial^{\lambda}\varphi_{00}$
can be ignored with respect to the ${\varphi^{00}}\partial_{\lambda}\varphi^{00}\partial^{\lambda}\varphi^{00}$
term: the terms $\left[\partial\varphi\partial\varphi\right]$ lead
to linear field equations, so they do not contribute to the non-linearity
due to the $[\varphi^{n}\partial\varphi\partial\varphi]$ terms, while
they remain negligible compared to the linear terms $\partial_{\lambda}\varphi^{00}\partial^{\lambda}\varphi^{00}$.
The $\partial_{\lambda}\varphi^{00}\partial^{\lambda}\varphi_{00}$
structure dominates all other components of the form $[\partial\varphi\partial\varphi]$.
Thus, if the influence of the ${\varphi^{00}}\partial_{\lambda}\varphi^{00}\partial^{\lambda}\varphi^{00}$
term is not negligible compared to the effect of the $\partial_{\lambda}\varphi^{00}\partial^{\lambda}\varphi^{00}$
term (i.e. there are strong field non-linearities) then it is justified
to neglect the components $\varphi^{\mu\nu}$ $(\mu\neq\nu)$ of $[\partial\varphi\partial\varphi]$.
The corrections coming from the modification of the Euler-Lagrange
equation enter only at the level of the terms $\varphi^{n+1}\partial^{2}\varphi$,
with $n\geq2$, and can likewise be neglected. Hence, in the static case --which 
identifies to the high-temperature limit in Euclidean space--, the GR Lagrangian may be simplified to

\begin{equation}
\mathcal{L}_{\mathrm{GR,stat.}}=\sum_{n=0}^{\infty}\left(16\pi G\right)^{n/2}\varphi^{n}\left[a_{n}\partial\varphi\partial\varphi-T_{00}\right],\label{eq:Scalar Lgrangian}
\end{equation}
where the subscript $\mathrm{\scriptsize{stat.}}$ reminds us that $\mathcal{L}{}_{\mathrm{GR}}$
is expressed in the static limit. We define the scalar field $\varphi\equiv\varphi_{00}$. We found above that $a_0 =1$
and confirm it in Appendix~\ref{Appendix: Spin 0 Assumption} by deriving the weak field potential from
Eq.~(\ref{eq:Scalar Lgrangian}), and comparing it to the Einstein\textendash{}Infeld\textendash{}Hoffmann
equations~\cite{Einstein Infled Hoffmann}. The values of the other $a_{n}$ coefficients are not important here since
Eq.~(\ref{eq:Scalar Lgrangian}) will be used only at first orders.  We set $a_{n}=1$ for all $n$ in the rest of the paper.  For
 two static point sources located at $x_{1}$ and $x{}_{2}$, $\mathcal{L}_{\mathrm{GR,stat.}}$
becomes
\begin{eqnarray}
\mathcal{L}_{\mathrm{GR,stat.}}=\sum_{n=0}^{\infty}\left(16\pi G\right)^{n/2}   \biggl[\varphi^{n}\partial\varphi\partial\varphi-
\left(16\pi G\right)^{1/2}\,\varphi^{n+1}\left(\delta^{(4)}(x-x_{1})+\delta^{(4)}(x-x_{2})\right)^{n+1}\biggr]. \label{eq:Scalar Lagrangian for two sources} 
\end{eqnarray}

Gravitation is a spin-2 theory and, like any spin-even theory, is always attractive. Satisfactorily, the scalar, spin-0, approximation is also spin-even and thus also always attractive. 
In all, we see that the pure field part of $\mathcal{L}_{\mathrm{GR,stat.}}$ is identical to Eq.~(\ref{eq:Toy Lagrangian GR}), thus supporting
that GR can be adequately described in its static limit with the Lagrangian in Eq.~(\ref{eq:Toy Lagrangian GR}).

Likewise in QCD, the chromoelectric component $A^a_0 \equiv A$ of the gluon field dominates in the case of two static sources and the QCD Lagrangian may be approximated by
\begin{eqnarray}
\mathcal{L}=\partial_{\mu}A\partial^{\mu}A-\pi \alpha_s A^{4}. 
%\\ \nonumber 
\label{eq:Toy Lagrangian QCD 2}
\end{eqnarray}

Contrary to Eq.~(\ref{eq:QCD Lagrangian}), Eq.~(\ref{eq:Toy Lagrangian QCD 2}) has no cubic term. 
In the strong coupling regime where non-perturbative effects are dominant, either $\pi \alpha_s A^4 \gg \partial A \partial A$ 
if only the quartic term dominates in Eq.~(\ref{eq:QCD Lagrangian}), 
or $\sqrt{\pi \alpha_s} A^2 \gg \partial A$ if the cubic term dominates over the quadratic 
one. The latter inequality also implies $\sqrt{\pi \alpha_s} A^4 \gg A^2 \partial A$ so in any case, the 
quartic term dominates over the other terms. Thus, the lack of a cubic term in Eq.~(\ref{eq:Toy Lagrangian QCD 2}) may 
not preclude that this equation describes QCD in a classical limit. 
In fact, it has been advanced that Eq.~(\ref{eq:Toy Lagrangian QCD 2}) describes QCD in
either the classical limit when quantum fluctuations are neglected, or in the low energy, strong coupling, limit ~\cite{Frasca:2007uz,Frasca:2009yp}.

In QCD, a two-source system corresponds to a meson which, to be colorless and have zero baryon 
number, must be made of constituent quark and antiquark of opposite colors. 
QCD being a spin-1 theory, the force either attracts or repulses. However, the 
force between the two static sources is solely attractive since they must carry opposite colors. Hence, as for gravity, the 
always attractive  spin-0 field is also satisfactory in the case of strong interaction. This also applies
to baryons when viewed as quark-diquark structures.%, but we will not consider three-sources system here.

In all, the QCD and GR Lagrangians, Eqs.~(\ref{eq:QCD Lagrangian}) and~(\ref{eq:Polynomial Einstein-Hilber Lagrangian}), respectively,
may be approximated in the classical and static limits by the scalar Lagrangians 
of Eqs.~(\ref{eq:Toy Lagrangian QCD 2}) and~(\ref{eq:Scalar Lgrangian}), respectively.
We remark that the invariance principles underlying the theories, e.g. gauge invariance 
under SU(3)$_{c}$ for QCD, cannot be considered in scalar theories.
However, they are manifest in the sense that the form of the scalar Lagrangians stem from the ones of the full theories whose 
Lagrangian forms themselves are imposed by invariance principles. 
Likewise, the non-abelian nature of the theories cannot be formalized with a scalar theory but is still manifest in 
the non-linear terms of the Lagrangian. Related features that the full theories possess, e.g. different colors charges, 
must be accounted for ad hoc, e.g. by inserting the appropriate 
Casimir factor in the static potential. 

\subsection{Onset of the strong regime for QCD and GR \label{sec: strong regime onset}}
The QCD coupling  at long distance is large~\cite{Deur:2016tte}. 
There, effects from the higher order cubic and quartic terms of
Eq.~(\ref{eq:QCD Lagrangian}) become dominant and quarks are confined. 
In the gravitational force case, what can lead to a non-perturbative regime is either interactions at very short distances 
(presumably described by quantum gravity, which we do not consider here), or  
a large system mass $M$, which is what concerns us here. The gravitational force 
is a function of its total field squared, $\varphi^2$ and is also proportional to $M$. Normalizing $\varphi$
to  $\varphi^{2}=M\phi^{2}$ makes explicit the importance of the higher order terms $\left(16\pi G\right)^{n/2}M^{n/2+1}\varphi^{n}\partial\varphi\partial\varphi$
in Eq.~(\ref{eq:Scalar Lagrangian for two sources}) for large values
of $M$. Rescaling $\mathcal{L}$
by $ $$1/M$, the three-dimensional action based on Eq.~(\ref{eq:Scalar Lagrangian for two sources})
is 
\begin{eqnarray}
S=\int\mathrm{d}^{3}x\bigl[\partial_{i}\phi\partial^{i}\phi+g\phi\partial_{i}\phi\partial^{i}\phi+g^{2}\phi^{2}\partial_{i}\phi\partial^{i}\phi+\cdots \label{eq:Full Action GR} 
-g\phi\left(\delta^{(3)}(x-x_{1})+\delta^{(3)}(x-x_{2})\right)-\cdots\bigr],\end{eqnarray}
with $g=\sqrt{16\pi MG}$.
(The integration of the Lagrangian density being here three-dimensional; see Sect.~\ref{sub:Quantum-vs-classical},
we summed over the spatial index $i$ rather than the Lorentz index $\mu$.)
We note that rescaling thus the Lagrangian emphasizes further the  quantum effect suppression
discussed in Sect.~\ref{sub:Quantum-vs-classical}. In all, quantum
effects for the theory given by Eq.~(\ref{eq:Scalar Lagrangian for two sources}) are suppressed by $\hbar/M\tau$,
with $M$ a large mass in our regime of interest.

 In QCD, the non-linear regime arises for distances greater than $2 \times 10^{-16}$ m when 
 $\alpha_s$ becomes large: $\alpha_s \simeq 0.5$ in the $\overline{MS}$ scheme~\cite{Deur:2016tte}. 
 Likewise, such a regime should arise in GR when
 $g=\sqrt{16 \pi G M}$ is large, with $M$ the system mass, and the system size is comparable to the field correlation length $L$. 
 For a galaxy, typically $\sqrt{16 \pi G M/L} \simeq 10^{-2}$, with $L=10$ kpc taken to be a third of a typical galaxy size,
 and $M=3 \times 10^{11} M_\odot$, a typical galaxy mass. 
$g \simeq 10^{-2}$ is large enough to trigger the non-linear regime onset; see Fig.~\ref{Flo:size effect}.

\subsection{Applying the scalar model to QCD and GR in the high-temperature limit  \label{sec:Interpretation GR}}
\subsubsection{QCD}
\begin{figure}[ht!]
\centering
 \includegraphics[width=0.6\textwidth]{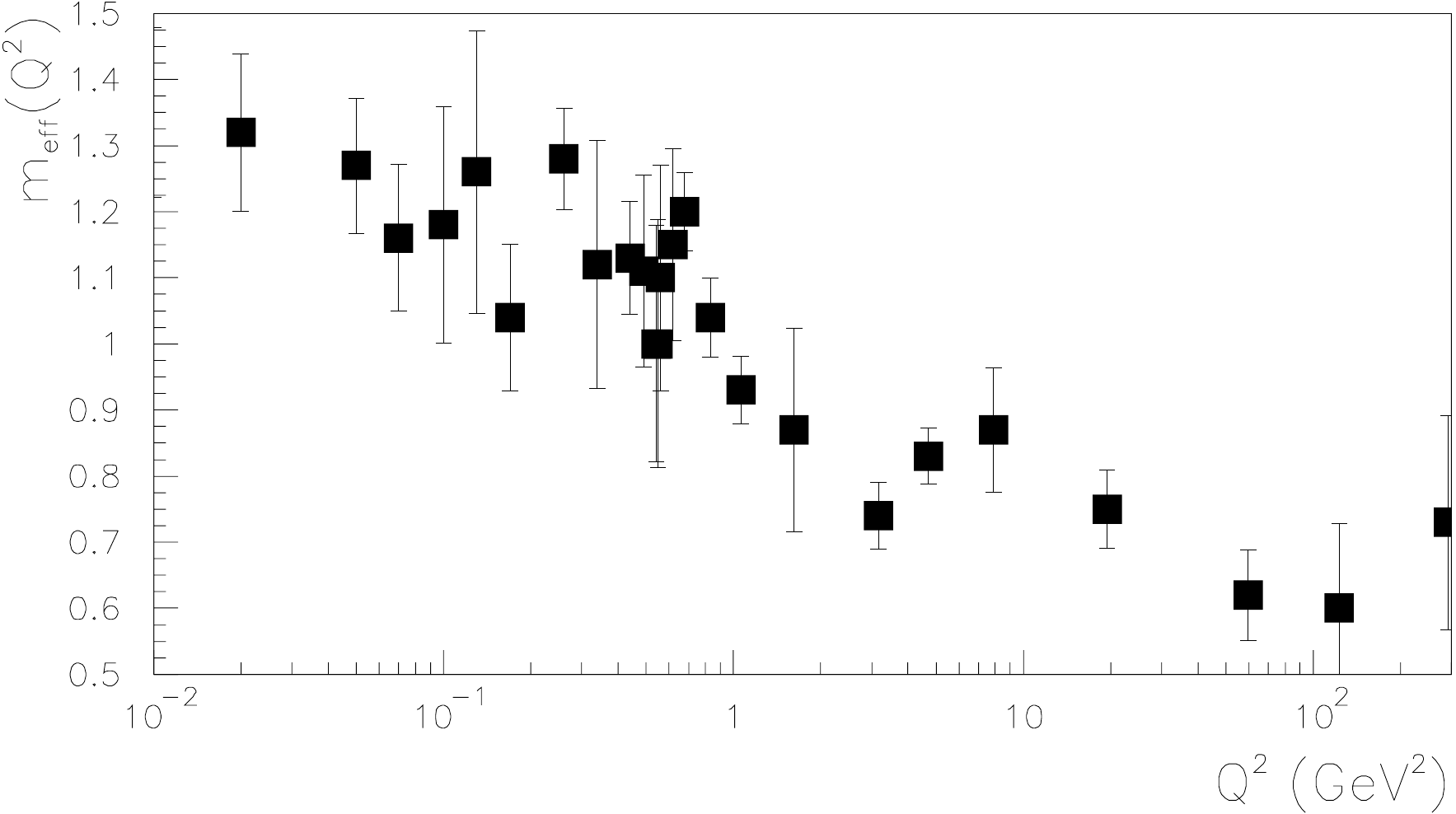}
\caption{
\label{fig:mass vs alpha_s2}The running effective mass $m_{\mbox{\scriptsize{eff}}}(Q^2)$ as a 
function of the 4-momentum transfer $Q^2$ corresponding to the $g^2$ values shown in 
Fig.~\ref{fig:mass vs alpha_s}
\protect \\
}
\end{figure}

While GR is a classical theory, and thus relevant for comparison with results obtained using the Lagrangian in 
Eq.~(\ref{eq:Scalar Lagrangian for two sources}), there is \emph{a priori} no well-known  classical limit  
pertinent to QCD, although semi-classical approximations do exist~\cite{Brodsky:2014yha,Hoyer:2009ep,Dietrich:2012un}. 
Certainly, an approximate Yukawa potential such as the one obtained in the high-temperature limit for the $g^2 \phi^4$ theory;
see Fig.~\ref{Flo:Boundary conditions, qcd case}, does not resemble the linear potential expected in the QCD static case. This
indicates that, provided  Eq.~(\ref{eq:Toy Lagrangian QCD}) indeed describes the classical aspect of QCD,
quantum effects are essential for obtaining a linear potential.  
Whereas, in the pure field sector of QCD, quantum effects such as quark 
condensates are irrelevant, others such as the running of the coupling cannot be disregarded when comparing meaningfully with phenomenology. 
Accounting for it can be done by mapping the $ g^2$-dependence of $m_{\mbox{\scriptsize{eff}}}$ into a 4-momentum 
transfer $Q^2$-dependence assuming that $g^2/\pi$ runs as the QCD coupling $\alpha_s(Q^2)$~\cite{Deur:2016tte}.
(Equations~(\ref{eq:Toy Lagrangian QCD}) and~(\ref{eq:QCD Lagrangian}) allow us to identify $g^2/\pi$ to $\alpha_s$.)
By augmenting the $g^2 \phi^4$ theory with the running of the coupling, we introduce short-distance quantum
effects in an otherwise classical setting. 
We used the AdS/QCD results on $\alpha_s(Q^2)$~\cite{Brodsky:2010ur,Deur:2014qfa,Deur:2016cxb} for 
$Q^2 \lesssim 1$ GeV$^2$, supplemented by the perturbative
QCD expression of  $\alpha_s(Q^2)$ at fourth order for larger $Q^2$. 
Such $\alpha_s(Q^2)$  agrees well with the phenomenology~\cite{Deur:2005cf,Deur:2008rf} for all $Q^2$.
Specifically, we used $\alpha_s(Q^2)$ in the conventional $\overline{MS}$ renormalization scheme. 
The result is shown in Fig.~\ref{fig:mass vs alpha_s2}.

\begin{figure}[ht!]
\centering
\includegraphics[width=0.6\textwidth]{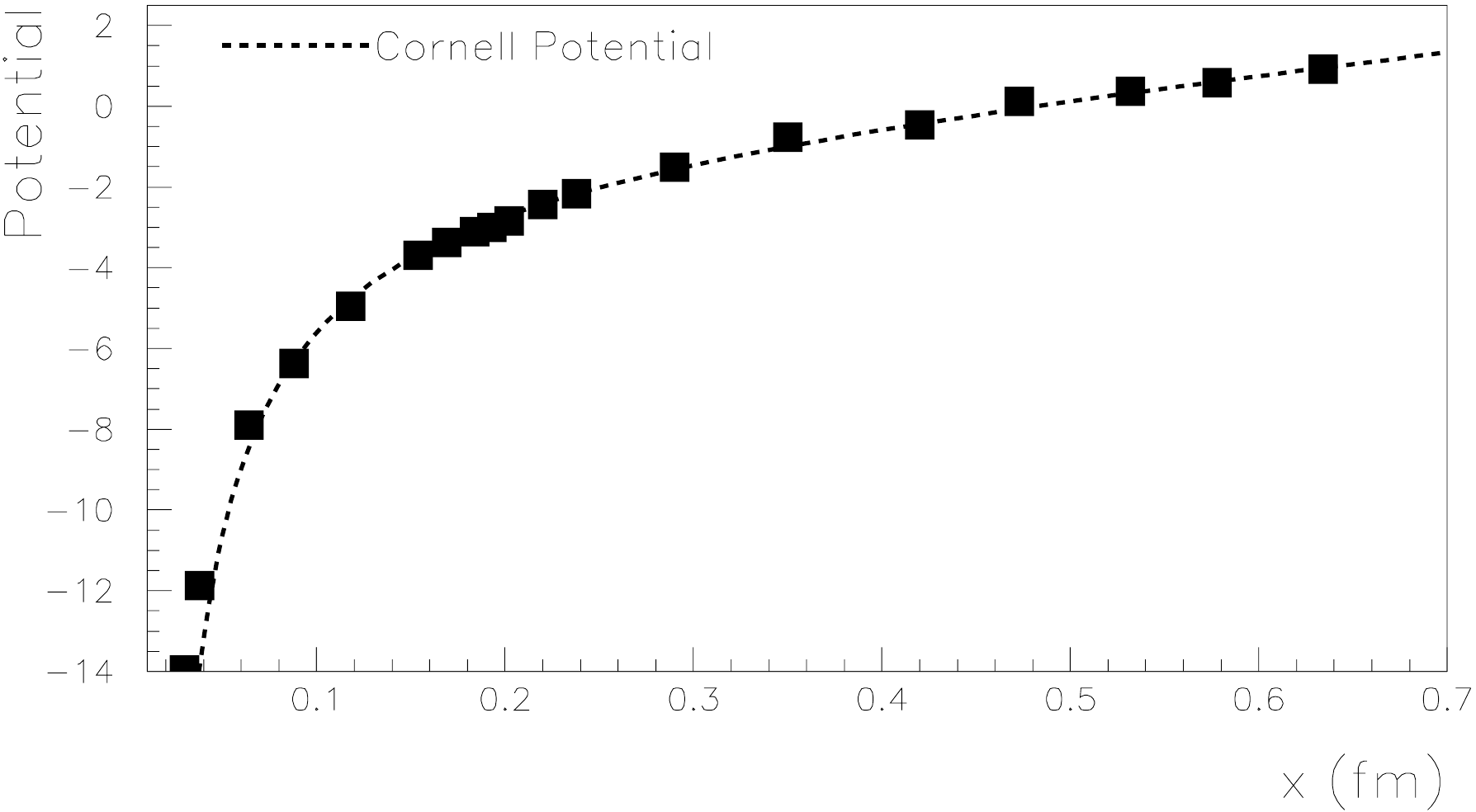}
\caption{
\label{fig:QCD linear pot}High-temperature potential from the $g(x)^2 \phi^4$ theory as a function of distance after adding 
the runnings of $g^2$ and of $m_{\mbox{\scriptsize{eff}}}$. The strong interaction's Cornell potential is shown by the \emph{dashed line}
\protect \\
}
\end{figure}

Using these results, the potential of the $g^2(x) \phi^4$ theory becomes
\begin{equation}
G_{2p}(x) = -C_f  \frac{g^2(x)}{\pi} \frac{e^{-m_{\mbox{\scriptsize{eff}}}(x)x}}{x}, \label {QCD V(x)}
\end{equation}
where we included the appropriate color factor $C_f =4/3$ for a more pertinent comparison with QCD, and where $g^2(Q^2)$ and 
$m_{\mbox{\scriptsize{eff}}}(Q^2)$ were Fourier-transformed to position space. The result 
is shown in Fig.~\ref{fig:QCD linear pot} and compared with the well-established Cornell 
potential~\cite{Eichten:1974af,Eichten:1978tg,Eichten:2002qv} that successfully 
describes hadron spectroscopy and that is consistent with quenched SU(3) QCD lattice 
calculations~\cite{the:Bali 1995 et al.}. $G_{2p}(x)$ displays the expected linear 
behavior in the relevant $x$ range and agrees well with the Cornell potential. 
(For this comparison,  $x$ in Eq.~(\ref{QCD V(x)})  must be rescaled by $1/C_f$, since not having color factors
in Eq.~(\ref{eq:Toy Lagrangian QCD 2}) underestimates the magnitude of the field coupling by $C_f$ compared to the SU(3) QCD theory.)
The string tension, fit from the results between 0.35 and 0.6 fm in Fig.~\ref{fig:QCD linear pot}, is $\sigma=0.157 \pm 0.012$ GeV$^2$.
This yields a confinement parameter $\kappa = \sqrt{ \sigma\pi/2} = 0.496 \pm 0.038$ GeV that agrees with the value 
$\kappa =  0.523 \pm 0.024$ GeV found in~\cite{Brodsky:2016yod}.

The appearance of a mass scale in the $g^2 \phi^4$ theory, despite its conformal Lagrangian, is a valuable feature.  
Understanding the emergence of a GeV-size mass scale is one of the outstanding problems of strong Interaction; see 
e.g.~\cite{Brodsky:2014yha}. In the present work, such emergence can be understood from the absence of derivatives 
in the quartic term $g^2 \phi^4$, which thus does not contribute to field propagation but acts as a field mass. 
QCD also possesses a non-propagating $A^a_\mu A^b_\nu A^{\mu c} A^{\nu d}$ term in its Lagrangian. 
This suggests that the mass scale emergence in strong Interaction is a classical effect due to the specific 
form of the quartic term of the QCD Lagrangian. The cubic term may contribute similarly, or may instead  generate a 
long correlation length as it does in the GR case. Although the good agreement of the $g^2(x) \phi^4$ theory
with strong Interaction suggests a minor role of the cubic term, its actual effect cannot be explored in the 
present model. 

In all, the running effective mass and the potential obtained from the Lagrangian in Eq.~(\ref{eq:Toy Lagrangian QCD 2}) 
are consistent with what is expected from QCD.
If this consistency underlines the pertinence of the $g(x)^2 \phi^4$ theory to model QCD, then it indicates that quantum 
effects, responsible for the running of the strong coupling, play a role in shaping the potential.
Augmenting the $g^2 \phi^4$ theory at high-temperature with the running of $g^2$ is
necessary to map the values of the field's effective mass into a $Q^2-$running mass, 
and to change the Yukawa potential into a linear potential.
(An equivalent solution leading to a linear potential without introducing a running of the coupling would be to 
express the dressed gluon propagator as a sum of massive field propagators as done in 
Refs.~\cite{Frasca:2007uz,Frasca:2009yp,Frasca:2015wsa}).
A consistency between the $g^2(x) \phi^4$ theory and QCD  also encourages 
the application of the present approach to the more complex case of GR, 
\emph{a fortiori} since GR is already a classical theory and would not need to be supplemented ad hoc by quantum effects.
\subsubsection{General Relativity}
The ($g \phi \partial_\mu \phi \partial^\mu \phi + g^2 \phi^2 \partial_\mu \phi \partial^\mu \phi$) theory
calculations in the high-temperature limit yield a potential which
varies approximately linearly with distance; see Figs.~\ref{Flo:size effect} and~\ref{Flo:size effect Newton subtracted}.
This can be pictured as a collapse of the three-dimensional
system into one dimension. As discussed in Sect.~\ref{sec: strong regime onset} and shown numerically, typical galaxy masses are 
enough to trigger the onset of the strong regime for GR. 

Hence, for two massive bodies, such as two galaxies or two galaxy clusters, this would result in a string containing a 
large gravity field that links the two bodies --as suggested by the map of the large 
structures of the universe. That this yields quantitatively the observed dark mass
of galaxy clusters and naturally explains the Bullet Cluster observation~\cite{Clowe:2006eq} was discussed in~\cite{Deur:2009ya}.

For a homogeneous disk, the potential becomes logarithmic. Furthermore, 
if the disk density falls exponentially with the radius, as it is the case for disk galaxies,
it is easy to show that a logarithmic potential yields flat rotation curves: a body subjected to such a 
potential and following a circular orbit (as stars do in disk galaxies to good approximation) follows the equilibrium equation:
\begin{equation}
v(r)=\sqrt{G'M(r)},\label{eq:frc1}
\end{equation}
with $v$ the tangential speed and $M(r)$ the disk mass integrated up to $r$,  the orbit radius. $G'$ is an effective
coupling constant of dimension GeV$^{-1}$ similar to the effective coupling $\sigma$ (string tension) in QCD. 
Disk galaxies density profiles typically fall exponentially:  $\rho(r)=M_{0}e^{-r/r_{0}}/(2\pi r_{0}^{2})$, where
$M_{0}$ is the total galactic mass and $r_{0}$ is a characteristic  length particular to a galaxy. 
Such a profile leads to, after integrating $\rho$ up to $r$:
\begin{equation}
v(r)=\sqrt{G'M_{0}\left(1-(r/r_{0}+1)e^{-r/r_{0}}\right)},\label{eq:frc2}
\end{equation}
At small $r$, the speed rises as $v(r)\simeq\sqrt{G'M_{0}}r/r_{0}$ and flatten at large $r$: $v\simeq\sqrt{G'M_{0}}$.
This is what is observed for disk galaxies and the present approach yields rotation 
curves agreeing quantitatively with observations~\cite{Deur:2009ya}. Flat rotation 
curves are an essential feature of galaxy dynamic that is natural in the present approach. In contrast,
they are implemented in an ad hoc way in the standard WIMP/axion approach to dark matter, the profile of the
dark matter halo being chosen specifically to yield flat rotation curves. 

A problematic observation for non-baryonic dark matter is the correlation between the rotation speeds 
and baryonic masses of disk galaxies, a phenomenon known as the Tully-Fisher relation~\cite{Tully:1977fu}. 
The relation is puzzling since it is a dynamical relation that does not involve the galaxy dark mass while this
one dominates the total galaxy mass. We showed in~\cite{Deur:2009ya} how this relation is also naturally
explained from the logarithmic potential arising in a disk galaxy. This mass--rotation relation of GR can be 
paralleled to the mass--angular-momentum relation of QCD at the origin of the Regge 
trajectories~\cite{Regge:1959mz,the: Greensite conf.}. The two relations are strikingly similar.

For a uniform and homogeneous spherical distribution
of matter, the system remains three-dimensional and the static potential
stays proportional to $1/r$, in contrast to the linear or logarithmic potentials of 
the one- and two-dimensional cases, respectfully.   
The dependence of the potential with the system's symmetry suggested a search for a
correlation between the shape of galaxies and their dark masses~\cite{Deur:2009ya}.
Evidence for such correlation has been found~\cite{Deur:2013baa}.

\section{Summary and conclusion}
We have numerically studied non-linearities in two particular scalar field theories. 
In the high-temperature limit these may  provide a possible description of the strong interaction 
and of General Relativity (GR) that would simplify
their study in their non-perturbative regime, and allow for fast numerical calculations.
The overall validity of the method is verified by recovering analytically known potentials. We further verified the validity
of the simplifications in the case of GR by recovering the post-Newtonian formalism.

Several approaches to QCD in its non-perturbative regime already exist, such as e.g. Lattice Gauge Theory, Dyson-Schwinger Equations,
Functional Renormalization Group, Stochastic Quantization, or the AdS/CFT correspondence to name a few. 
They are more advanced and more exact that the simplified method discussed here. 
What primarily justifies the development  of this simpler approximation to QCD is that it provides further 
checks of the method, which we can then be applied with confidence to the GR case. 
A subsidiary benefit of testing the method with QCD is that it may help to isolate important ingredients leading to confinement.
In that context, this method is able to provide: 
(1) the expected field function, 
(2) a mechanism --straightforwardly applicable to QCD-- for the emergence of a mass scale out of a conformal Lagrangian, 
(3) a running of the field effective mass, and 
(4) a potential agreeing with the phenomenological Cornell linear potential up to 0.8 fm, the relevant range for hadronic physics. 
Quantum effects, which cause couplings to run, are necessary for producing the linear potential
and must be supplemented to the approach. 
A non-running coupling, even with a large value, would only yield a short range Yukawa potential.
That  the method recovers the essential features of QCD supports its application to GR. Since GR is a classical theory, no
running coupling needs to be supplemented. The main goal of the paper is to provides a simple but reliable method
to study GR in its non-perturbative regime since compared to QCD, 
non-perturbative methods for GR are not as developed. It is interesting to
develop a method for GR that relates to QCD's formalism since QCD and GR share similar field Lagrangians,
with cubic and quartic field self-interactions terms, and since 
--contrary to GR-- QCD's non-perturbative phenomenology is well studied both
experimentally and theoretically. QCD's phenomenology may then serve as a guide 
to suggest, identify and understand non-perturbative effects in GR.

In fact, the similarity between QCD and GR field Lagrangians may explain
intriguing parallels between on the one hand observations interpreted with dark matter and on the other hand the phenomenology of 
the strong interaction. For examples, the total masses of galaxies and galaxy clusters are much 
larger than the sum of the masses of their known components, just like for hadrons; galaxies and hadrons share similar mass-rotation 
correlations (the Tully-Fisher relation and Regge trajectories, respectively); the large-scale balancing of 
gravitational attraction by dark energy --known as the cosmic coincidence problem-- may be paralleled with the 
large-distance suppression of the strong interaction into a much weaker Yukawa force; the universe's large 
scale stringy structure observed by weak gravitational lensing is evocative of  massive bodies, such as galaxies 
--or at larger scale two galaxy clusters-- linked by a QCD-like string/flux tube.

These parallels and the similar structure of GR and QCD's Lagrangians suggest that massive galactic objects such
as galaxies or cluster of galaxies have triggered the non-perturbative regime of GR and that dark matter and dark energy
are manifestations of it. It is clear that the self-interactions terms in GR have the right effects to explain dark matter and dark energy,
the question being whether galaxies are massive enough to trigger these effects. This is the question we addressed in the second
part of this paper.
In QCD, the non-perturbative regime arises for distances greater than $2 \times 10^{-16}$ m, while we find that for GR it arises at galactic scales. 
The non-perturbative effects then explain quantitatively the dynamics of spiral galaxies and clusters~\cite{Deur:2009ya}.
In particular, in the case spiral galaxies, the logarithmic potential 
resulting from the strong field non-linearities trivially yields flat rotation curves. Finally, for a uniform and homogeneous spherical distribution
of matter, the non-linearity effects should balance out~\cite{Deur:2009ya}. Evidences for the consequent correlation expected between galactic ellipticity 
and galactic dark mass  have been found~\cite{Deur:2013baa}. We thus have a natural explanation for dark matter and dark energy, which
according to the method developed and tested in this article, agrees with galactic and cluster dynamics. 
With the most probable phase-spaces for WIMPS and axions dark matter candidates now ruled out, 
it is important to consider the present explanation for the nature of Dark Matter and Dark Energy.

{\bf Acknowledgments}
The author thanks  S. J. Brodsky, C. Munoz-Camacho, F. X. Girod-Gard, P. Hoyer, J.-F. Mathiot, A. Sandorfi,
S. \v{S}irca, B. Terzi\'{c} and X. Zheng for useful discussions on the work reported here.

\appendix

\section{Method \label{sub:Method}}
To numerically compute $G_{2p}(x_{1}-x_{2})$, Eq.~(\ref{eq:2pt green}),
a cubic lattice of $N^{3}$ sites is used. Each site is associated
with a field value $\phi$. The set of $N^{3}$ values is called a
field configuration. These values should be such that they minimize
$S$, i.e. the field obeys the Euler-Lagrange equations of motion.
To determine these values,  a Wick rotation is first performed so that
$\mathrm{e}^{\mathrm{-i}S}\rightarrow\mathrm{e}^{-S}$. The action
$S$ is unchanged since in the static case, the Euclidean and
Minkowski actions are the same. One then proceeds iteratively, following
the Metropolis algorithm: for each of the $N^{3}$ sites, $S$ is
computed. Next, the value of $\phi$ at a site is shifted randomly
and the change in action $\Delta S$ is computed. If $\Delta \leq 0$,
the change tends to minimize the action. In this case, the new $\phi$
value is retained because it makes the field configuration closer
to what it must be in reality. If $\Delta S > 0$, the new $\phi$ value
is kept if $\mathrm{e}^{-\Delta S} > \zeta$, with $\zeta$ a random
number between 0 and 1, and rejected otherwise. While this operation
is being applied to all the sites, the new configuration of $\phi$
converges toward the configuration obeying the equations of motion.
That is, the probability distribution of the field configuration follows
$\mathrm{e}^{-S}$. The procedure is repeated enough times and the
results are averaged until they converge to the physical solution,
while the uncertainty stemming from the random method is minimized.
We remark that with this method $Z\equiv1$ in Eq.~(\ref{eq:2pt green}).

\section{Practical example of a numerical simulation\label{Appendix First-numerical-simulation}}
We provide here a practical example of static potential computation for the 
($g \phi \partial_\mu \phi \partial^\mu \phi + g^2 \phi^2 \partial_\mu \phi \partial^\mu \phi$) theory.
The $g^2 \phi^4$ case is simpler and follows the same method.

\subsection{First simulation\label{sub:First-simulation}}
\begin{enumerate}
\item The first step is to choose an initial configuration for $\phi$,
that is to choose $N^{3}$ values of $\phi$, one for each of the
$N^{3}$ lattice sites. The $\phi$ can be set to 0 or to random values 
$-\varepsilon < \phi < +\varepsilon$.
The optimal value for $\varepsilon$, about 1, will be discussed in
the next step. For a simple first simulation, the values of $\phi$
at the boundaries can be set to 0 (see Appendix~\ref{Appendix boundary conditions}
for a discussion). A small $N$ value, e.g. $N=11$, can be used initially
(it can be increased once the program is debugged and optimized). 
\item The initial configuration is arbitrary. Before spending CPU time
to compute $G_{2p}(x_{1}-x_{2})$, one must {}``thermalize'' the
configuration, i.e. modify it so that it is already close to the
physical configuration. To do this, the configuration is updated $10\times N_{\mathrm{cor}}$
times (take $N_{\mathrm{cor}}=20$) except at the boundaries where
$\phi$ stays 0. Each update follows the Metropolis procedure discussed
in Sect.~\ref{sub:Method}. For each update of the configuration,
the $(N-2)^{3}$ $\phi$ are offset by random numbers distributed
uniformly between $\pm\varepsilon$. For debugging ease at this stage,
the lattice version of the action; see e.g. Eq.~(\ref{eq:numerical action})
for the ($g \phi \partial_\mu \phi \partial^\mu \phi + g^2 \phi^2 \partial_\mu \phi \partial^\mu \phi$) theory,
should be used. The configuration updates can subsequently be sped up considerably
by using only the local change in the action; see Eq.~(\ref{eq:local action}).
Once Eq.~(\ref{eq:local action}) is used, $N$ can be increased to
25.
\item To further speed-up the simulation and minimize configuration correlations,
$\varepsilon$ must be optimized. The optimal value for $\varepsilon$
is such that during updates, about half of the $\phi$ values are changed once the
configuration is thermalized. 
\item One can now compute $G_{2p}(x_{1}-x_{2})$. For simplicity, we can
choose $x_{1}=(0,0,0)$, i.e. $x_{1}$ is the center of the lattice, choose
$x_{2}$ along a single direction, say $x$, and stop just before
the boundary. In that case, $G_{2p}(\pm n)$=$\phi(0,0,0)\phi(\pm na,0,0)$,
with $1<n<(N-3)/2$, the number of possible distances. The computations
are repeated $N_{\mathrm{s}}$ times and averaged for each given distance
$n$. Use $N_{\mathrm{s}}=5000$. To suppress correlations, between
each computation of $G_{2p}$, the configuration is updated $N_{\mathrm{cor}}$
times following step 2. (This is necessary because two configurations
differing by only one update are still similar, hence correlating
the two $G_{2p}$ and biasing the statistical averaging). 
\end{enumerate}
For a simulation using Eq.~(\ref{eq:local action}), with $N^{3}=(25)^{3}$, $N_{\mathrm{cor}}=20$, $N_{\mathrm{s}}=5000$
and $\varepsilon=1.2$, a typical simulation takes approximately 3
min on a MacBook Pro equipped with a 2.8 GHz Intel Core 2 Duo processor. 
To assess how this compares to other numerical approaches one can extend this
figure to state-of-art, 10 pflops, lattice machines. We can also take advantage of the fact that,
once the boundary conditions discussed in Appendix~\ref{Appendix boundary conditions}
are implemented, the Green function can be computed in any possible direction
at essentially no extra CPU-cost. In that case the relative precision of the calculation
would be $2 \times 10^{-7}$ for a 1--min calculation, comparable to the precision of a 32-bit machine. 

\subsection{Refinements}
\subsubsection{Faster updating\label{sub:Faster-updating}}
\begin{enumerate}
\item To update a configuration, the discretized expression of the action
is needed. The action below is the same as Eq.~(\ref{eq:Full Action GR})
but ignoring term of order $g^{3}$ or higher. Also, a 
field mass $m$ is added for checking purposes.  There is one source located at $x_{1}$: 
\begin{eqnarray}
S=\frac{1}{2}\int\mathrm{d}^{3}\phi\,\bigl[{(\partial_{i}\phi\partial^{i}\phi+}g\phi^{2}\partial_{i}\phi\partial^{i}\phi \label{eq:Lagrangian 3} 
+g^{2}\phi\partial_{i}\phi\partial^{i}\phi-g\phi\delta(x_{1})+m^{2}\phi^{2})\bigr].
\end{eqnarray}
 The $\phi^{n}\partial\phi\partial\phi$ terms can be transformed
into $\frac{1}{n+1}$$\phi^{n+1}\partial^{2}\phi$ by integrating
by parts. Recalling that, numerically, $\partial^{2}\phi(x)\approx\left[\phi(x+a)-2\phi(x)+\phi(x-a)\right]/a^{2}$,
with $a$ the distance between adjacent nodes, the numerical version
of the continuous action, Eq.~(\ref{eq:Lagrangian 3}) is 
\begin{eqnarray}
S=-\frac{1}{2}\sum_{n_{x}=2,n_{y}=2,n_{z}=2}^{N-1,N-1,N-1} \biggl(\bigl[\phi(n_{x}+1,n_{y},n_{z})+ 
\phi(n_{x}-1,n_{y},n_{z})+\phi(n_{x},n_{y}+1,n_{z}) \nonumber \\
+\phi(n_{x},n_{y}-1,n_{z}) +\phi(n_{x},n_{y},n_{z}+1)+\phi(n_{x},n_{y},n_{z}-1)-6\phi(n_{x},n_{y},n_{z})\bigr]
{}\bigl[\phi(n_{x},n_{y},n_{z}) \nonumber \\
+g\phi^{2}(n_{x},n_{y},n_{z})+g^{2}\phi^{3}(n_{x},n_{y},n_{z})\bigr]+g\phi(n_{x},n_{y},n_{z})\delta_{x_{1},y_{1},z_{1}} -m^{2}\phi^{2}(n_{x},n_{y},n_{z})\biggl),\label{eq:numerical action}
\end{eqnarray}
where $n_{x}$, $n_{y}$ and $n_{z}$ are the site indices in the
$x,y$ and $z$ directions, respectively, and the distance between
consecutive nodes has been set to $a=1$ (in lattice units).
\item Using Eq.~(\ref{eq:numerical action}) directly makes the simulation
computationally inefficient. However, since only the difference between
the actions before and after an update is used, we do not need to
sum the action over all the sites: only local terms near the site
($n_{x},n_{y},n_{z}$) that do not cancel in the difference are important.
The difference of the actions is \[
S_{_{\mathrm{new}}}-S_{_{\mathrm{old}}}\equiv\Delta S=-\frac{1}{2}\biggl(\bigl[\phi_{_{\mathrm{old}}}(n_{x}+1,n_{y},n_{z})+\]
\[
\phi_{_{\mathrm{old}}}(n_{x},n_{y}+1,n_{z})+\phi_{_{\mathrm{old}}}(n_{x},n_{y},n_{z}+1)-6\phi_{_{\mathrm{new}}}(n_{x},n_{y},n_{z})\]
\[
+\phi_{_{\mathrm{old}}}(n_{x}-1,n_{y},n_{z})+\phi_{_{\mathrm{old}}}(n_{x},n_{y}-1,n_{z})+\phi_{_{\mathrm{old}}}(n_{x},n_{y},n_{z}-1)\bigr]\]
\[
\bigl[\phi_{_{\mathrm{new}}}(n_{x},n_{y},n_{z})+g\phi_{_{\mathrm{new}}}^{2}(n_{x},n_{y},n_{z})+g^{2}\phi_{_{\mathrm{new}}}^{3}(n_{x},n_{y},n_{z})\bigr]\]
\[
-\bigl[\phi_{_{\mathrm{old}}}(n_{x}+1,n_{y},n_{z})+\phi_{_{\mathrm{old}}}(n_{x},n_{y}+1,n_{z})+\phi_{_{\mathrm{old}}}(n_{x},n_{y},n_{z}+1)\]
\[
-6\phi_{_{\mathrm{old}}}(n_{x},n_{y},n_{z})+\phi_{_{\mathrm{old}}}(n_{x}-1,n_{y},n_{z})+\phi_{_{\mathrm{old}}}(n_{x},n_{y}-1,n_{z})\]
\[
+\phi_{_{\mathrm{old}}}(n_{x},n_{y},n_{z}-1)\bigr] \bigl[\phi_{_{\mathrm{old}}}(n_{x},n_{y},n_{z})+g\phi_{_{\mathrm{old}}}^{2}(n_{x},n_{y},n_{z})\]
\[
+g^{2}\phi_{_{\mathrm{old}}}^{3}(n_{x},n_{y},n_{z})\bigr]
+\bigl[\phi_{_{\mathrm{new}}}(n_{x},n_{y},n_{z})-\phi_{_{\mathrm{old}}}(n_{x},n_{y},n_{z})\bigr]\]
\[
\bigl[\phi_{_{\mathrm{old}}}(n_{x}+1,n_{y},n_{z})+\phi_{_{\mathrm{old}}}(n_{x},n_{y}+1,n_{z})+\phi_{_{\mathrm{old}}}(n_{x},n_{y},n_{z}+1)\]
\[+\phi_{_{\mathrm{old}}}(n_{x}-1,n_{y},n_{z})+\phi_{_{\mathrm{old}}}(n_{x},n_{y}-1,n_{z})+\phi_{_{\mathrm{old}}}(n_{x},n_{y},n_{z}-1)\]
\[
+g[\phi_{_{\mathrm{old}}}^{2}(n_{x}+1,n_{y},n_{z})+\phi_{_{\mathrm{old}}}^{2}(n_{x},n_{y}+1,n_{z})\]
\[
+\phi_{_{\mathrm{old}}}^{2}(n_{x},n_{y},n_{z}+1)
+\phi_{_{\mathrm{old}}}^{2}(n_{x}-1,n_{y},n_{z})+\phi_{_{\mathrm{old}}}^{2}(n_{x},n_{y}-1,n_{z})\]
\[
+\phi_{_{\mathrm{old}}}^{2}(n_{x},n_{y},n_{z}-1)\bigr]
+g^{2}\bigl[\phi_{_{\mathrm{old}}}^{3}(n_{x}+1,n_{y},n_{z})\]
\[
+\phi_{_{\mathrm{old}}}^{3}(n_{x},n_{y}+1,n_{z})+\phi_{_{\mathrm{old}}}^{3}(n_{x},n_{y},n_{z}+1)+\phi_{_{\mathrm{old}}}^{3}(n_{x}-1,n_{y},n_{z})\]
\[
+\phi_{_{\mathrm{old}}}^{3}(n_{x},n_{y}-1,n_{z})+\phi_{_{\mathrm{old}}}^{3}(n_{x},n_{y},n_{z}-1)\bigr]\]
\[+g\bigl[\phi_{_{\mathrm{new}}}(n_{x},n_{y},n_{z})-\phi_{_{\mathrm{old}}}(n_{x},n_{y},n_{z})\bigr]\delta_{x_{1},y_{1},z_{1}}\]
\vspace{-0.5cm}
\begin{equation}
-m^{2}\bigl[\phi_{_{\mathrm{new}}}^{2}(n_{x},n_{y},n_{z})-\phi_{_{\mathrm{old}}}^{2}(n_{x},n_{y},n_{z})\bigr]\biggr),\label{eq:local action}
\end{equation}
\noindent where $\phi_{_{\mathrm{old}}}$ and $\phi_{_{\mathrm{new}}}$ denote the fields before and after update, respectively. 
\end{enumerate}

\subsubsection{Statistics increase }
To gain statistics at low CPU cost, $G_{2p}$ can be computed along
the y and z axes as well. To further gain statistics, it can also
be computed in more directions than just the three orthogonal axes.
However, in that case, boundary conditions $\phi=0$ on a sphere of
radius $N/2$ must be imposed. Otherwise, with the cubic zero boundary,
any strong $\phi$ autocorrelation would suppress $G_{2p}$ calculated
along one of the axis compared to a $G_{2p}$ calculated along e.g.
a diagonal.

\section{\label{Appendix boundary conditions} Boundary conditions \label{sub:Boundary-condition} }
The two-point Green function $G_{2p}(x)$ quantifies the field autocorrelation.
Weak correlations imply that the field values in most of the lattice
volume are independent of their values at the boundary. In that case
the choice of boundary conditions is not critical: see Fig.~\ref{Flo:Boundary conditions, Abelian case}
where $G_{2p}$ is calculated for a two-source system with free-fields. One can use periodic boundary conditions,
Dirichlet boundary conditions ($\phi=0$ at the boundaries), or random
values within the range $-\varepsilon < \phi <+\varepsilon$ ($\varepsilon$
is the optimal range within which $\phi$ is randomly shifted; see
Appendix~\ref{sub:First-simulation}).  
Likewise, in Fig.~\ref{Flo:Boundary conditions, qcd case} $G_{2p}$ is computed for different boundary conditions
in the $g^2 \phi^4$ case, using the Lagrangian in Eq.~(\ref{eq:Toy Lagrangian QCD}). The quartic term 
generates an effective mass. Thus the field correlation length is shortened and the
choice of boundary condition is even less important. We remark that, although Eq.~(\ref{eq:Toy Lagrangian QCD})
contains a non-linear term, i.e. a term involving field self-interactions, the total potential is well represented 
by the sum of the two individual potentials, thus obeying the field superposition principle. This is because there is 
no cubic term due to the equation of motion, and because the quartic term essentially acts as a  mass term. 

\begin{figure}[ht!] 
\centering
 \includegraphics[width=0.6\textwidth]{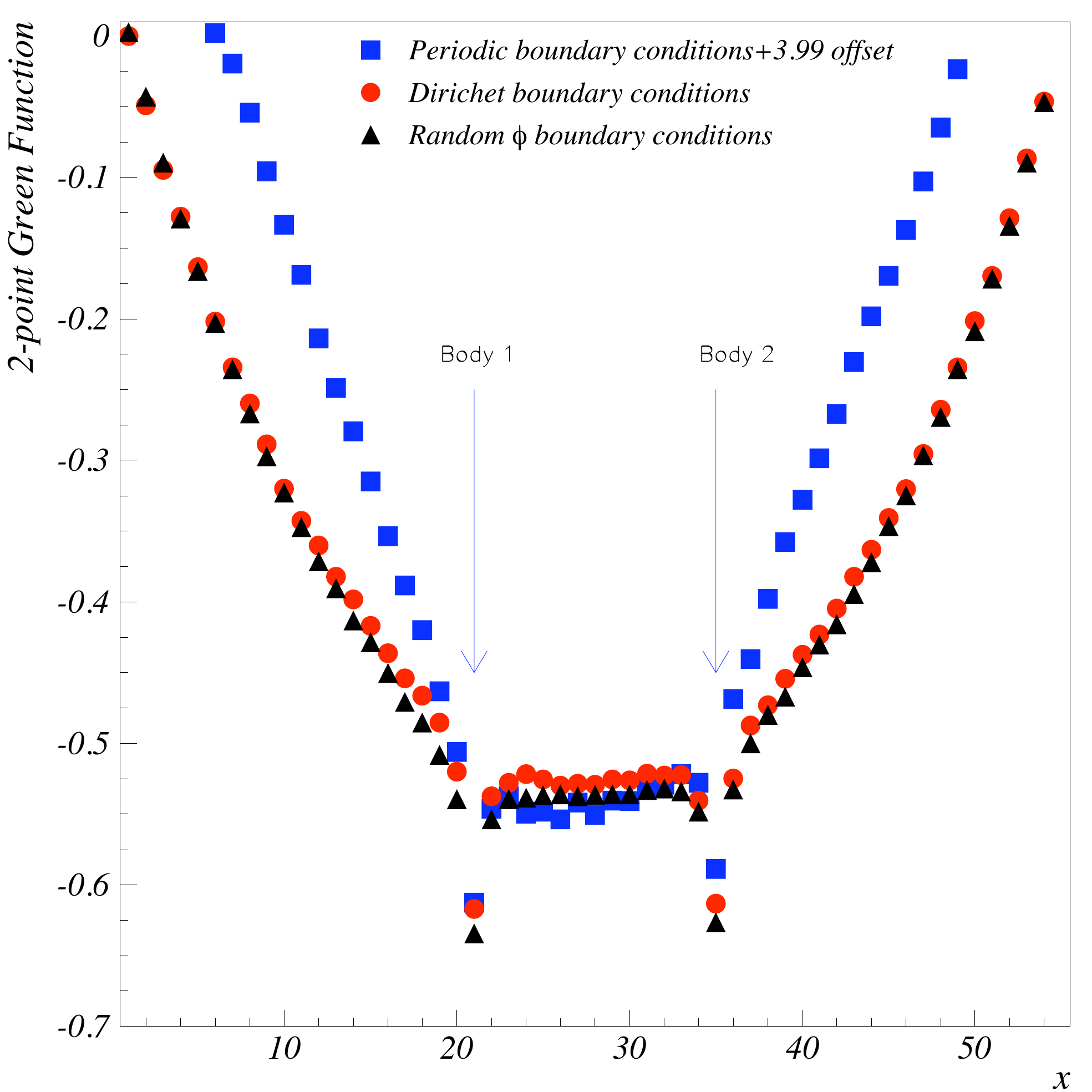}
\caption{
\label{Flo:Boundary conditions check} Same as Fig.~\ref{Flo:Boundary conditions, Abelian case} but using the full 
($g \phi \partial_\mu \phi \partial^\mu \phi + g^2 \phi^2 \partial_\mu \phi \partial^\mu \phi$) theory
Lagrangian, Eq.~(\ref{eq:Full Action GR}). The results are
essentially the same for Dirichlet and random field boundary conditions.
They are qualitatively similar for periodic boundary conditions but
quantitatively, the force between the two sources differs by up to a factor
of two. The cause of this difference originates from spurious effects in the
periodic boundary case due to the long correlation length allied with
the infinite repetitions of the sources
\protect \\
}
\end{figure}

In contrast, strong field autocorrelations
imply that the field at any position on the lattice could strongly
depend on the boundary values. The correlation length becomes 
an additional relevant physical scale to the problem. 
This is the case with the ($g \phi \partial_\mu \phi \partial^\mu \phi + g^2 \phi^2 \partial_\mu \phi \partial^\mu \phi$) 
theory, where covariant derivatives are present in each of the Lagrangian terms
and thus increase the field autocorrelation when 
$g$ becomes large. 
If the correlation length is sufficiently large and becomes comparable to
the lattice size then $\phi$ on any lattice site will depend on the choice
of the boundary conditions. In that case, $G_{2p}$ acquires
an unphysical dependence on the lattice size $N$, or equivalently on the lattice spacing for
a given system size. This unphysical phenomenon can be suppressed
by using large lattices made of an inner volume where the full action is used, and an outer volume where
we set $g=0$ (free-field case). The outer region provides a transition
buffer between the system and the arbitrary $\phi-$values at the
boundaries. This suppresses the dependence on the choice of the field
at the boundary and the $N-$dependence; see Figs.~\ref{Flo:Boundary conditions check}
and ~\ref{Flo: system size dependence}
where the Lagrangian of Eq.~(\ref{eq:Full Action GR}) was used. The results using the random
and Dirichlet boundary conditions agree well. The results with periodic
boundary condition are similar qualitatively but differ quantitatively.
The force computed under these conditions is about a factor of two larger
than with the random and Dirichlet boundary conditions. It is expected
that for correlation lengths comparable to or larger than the lattice
size, the repeated lattice pattern of the image sources spuriously
influences the field autocorrelation function anywhere in the lattice,
which is likely the cause for this difference between the result using the periodic
boundary conditions and the results using the two other conditions. 
A similar effect  is discussed in the context of linear potential in~Refs.~\cite{Deur:2009ya,Hoyer:2009ep}.
Another example of field with long correlation length is discussed
in Appendix~\ref{sub:Potentials-in-2-dimensional space}
where we recover the known $1/x$ free massless field potential in two-dimensional
space. This logarithmic potential represents fields that are significantly
autocorrelated and exemplifies the need for a careful handling of
boundary conditions to avoid spurious boundary effects. 
\begin{figure}[ht!] 
\centering
 \includegraphics[width=0.6\textwidth]{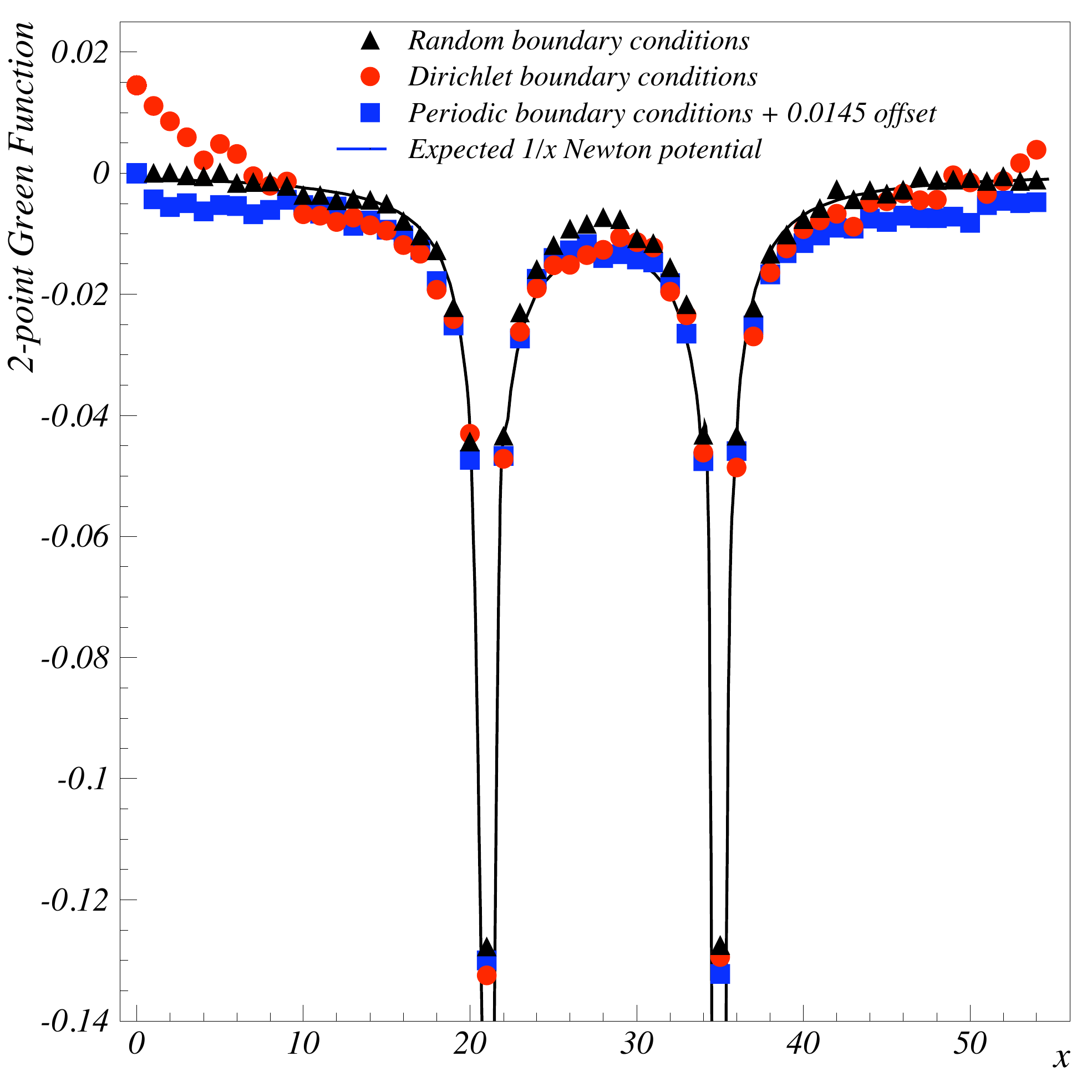}
\caption{
\label{Flo:Boundary conditions, Abelian case} Independence
of the simulation on the choice of boundary conditions. The results
are obtained when keeping only $\partial\phi\partial\phi$ and the
matter terms in the Lagrangian ($1/x$ free massless field case). The potential
$G_{2p}$ is plotted as a function of distance. The simulations were
run for a lattice size $N=55$, a coupling value $g=10^{-2}$, a correlation
coefficient (see Sect.~\ref{sub:First-simulation}) $N_{\mathrm{cor}}=20$
and for $N_{\mathrm{s}}=3.5\times10^{4}$ configurations. Two sources are
located on the $x-$axis at $d=\pm7$ from the lattice center at $x_{\mathrm{center}}=28$.
The \emph{circles} are for Dirichlet boundary conditions. The  \emph{squares}
are obtained with periodic boundary conditions. The  \emph{triangles}
are for random field boundary conditions. Except near the boundaries and
except for a constant offset of 0.0145 added to the result obtained
with the Dirichlet conditions (irrelevant since $G_{2p}$ is the potential),
all results agree  and display the expected $1/x$ potential
for each sources (\emph{continuous curve}) \protect \\
}
\end{figure}

\section{\label{Appendix verification} Method validation}
\subsection{Recovering analytically known potentials }
The method can be validated by verifying that analytically known
potentials are recovered. In this section we compute the free massless field,
free massive field (Yukawa) and $g^2 \phi^4$ theory static potentials in the case of three-dimensions. We also 
compute the massless and massive free-field static potentials in
two-dimensional space. We then compare them to their expected analytical
forms.

\subsubsection{Potentials in three-dimensional space}
\begin{figure}[ht!]
\centering
 \includegraphics[width=0.8\textwidth]{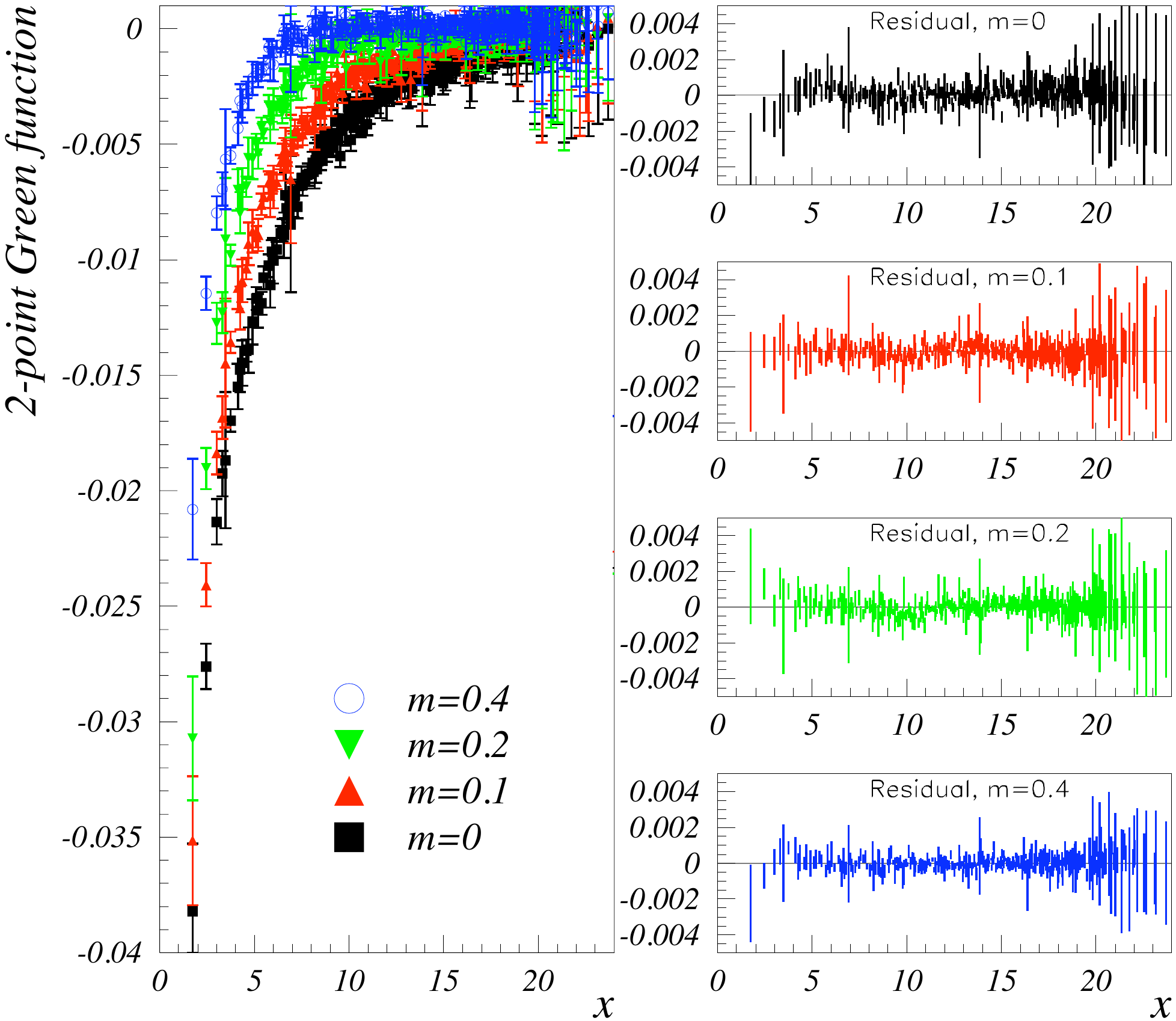}
\caption{
\label{fig: abelian case}
 Potentials obtained for the free (massless and massive) field cases
(\emph{left}). The free massless field potential is shown by the \emph{squares} ($m=0$).
The free massive field (Yukawa) potentials are obtained for different values of the field mass,
$m=0.1$ (\emph{up triangles}), $m=0.2$, (\emph{down triangles}) and
$m=0.4$ (\emph{open  circles}). Residuals from the $\mathrm{e}^{-mx}/x$
expectation are shown on the \emph{right}\protect \\
}
\end{figure}

In the massless free-field case, one should
recover a $G_{2p}(x)\propto1/x$  potential. Furthermore,
if one adds to the Lagrangian another quadratic term corresponding to a
field mass term, $m^{2}\phi^{2}$, one should recover a
Yukawa potential, $G_{2p}(x)\propto\mathrm{e}^{-mx}/x$. Such simulations
were ran using a lattice size $N=36$, $N_{\mathrm{s}}=10^{4}$ configurations,
a decorrelation coefficient (see Sect.~\ref{sub:First-simulation})
$N_{\mathrm{cor}}=10$ and random field boundary conditions. The results
are shown in Fig.~\ref{fig: abelian case}. The simulation agrees
well the expected potentials in both the free massless field and the Yukawa cases.

As discussed in the main text in  Sect.~\ref{sec:Interpretation QCD} the $g^2 \phi^4$ case is analytically known~\cite{Frasca:2009bc}. 
We recovered potentials well fit by a Yukawa potential or by the convolution of the Jacobi elliptic function $\mbox{sn}(x,-1)$, the expected function 
describing $\phi$~\cite{Frasca:2009bc}. The effective field mass stemming from $g^2 \phi^4$, shown in function of $Q^2$ in 
Fig.~\ref{fig:mass vs alpha_s2} qualitatively agrees with typical QCD expectation. 
Injecting the running field mass in the Yukawa potential form and Fourier transforming to position space yield the static potential 
shown in Fig.~\ref{fig:QCD linear pot}. It displays a linear behavior in the $x$--range relevant to hadron physics and agrees well with the Cornell 
potential. In all, the $g(x)^2 \phi^4$ theory recovers the principal confinement features expected from strong Interaction. 
If $g^2 \phi^4$ is the classical limit of QCD as proposed in~\cite{Frasca:2009yp}, the good match between the 
$g(x)^2 \phi^4$ theory and Cornell potentials suggests that 
the missing quantum effects can be fully encapsulated in the running of the coupling and that the method discussed here does provide relevant
approximations to QCD and GR.

\subsubsection{Potentials in two-dimensional space\label{sub:Potentials-in-2-dimensional space}}
\begin{figure}[ht!] 
\centering
 \includegraphics[width=0.6\textwidth]{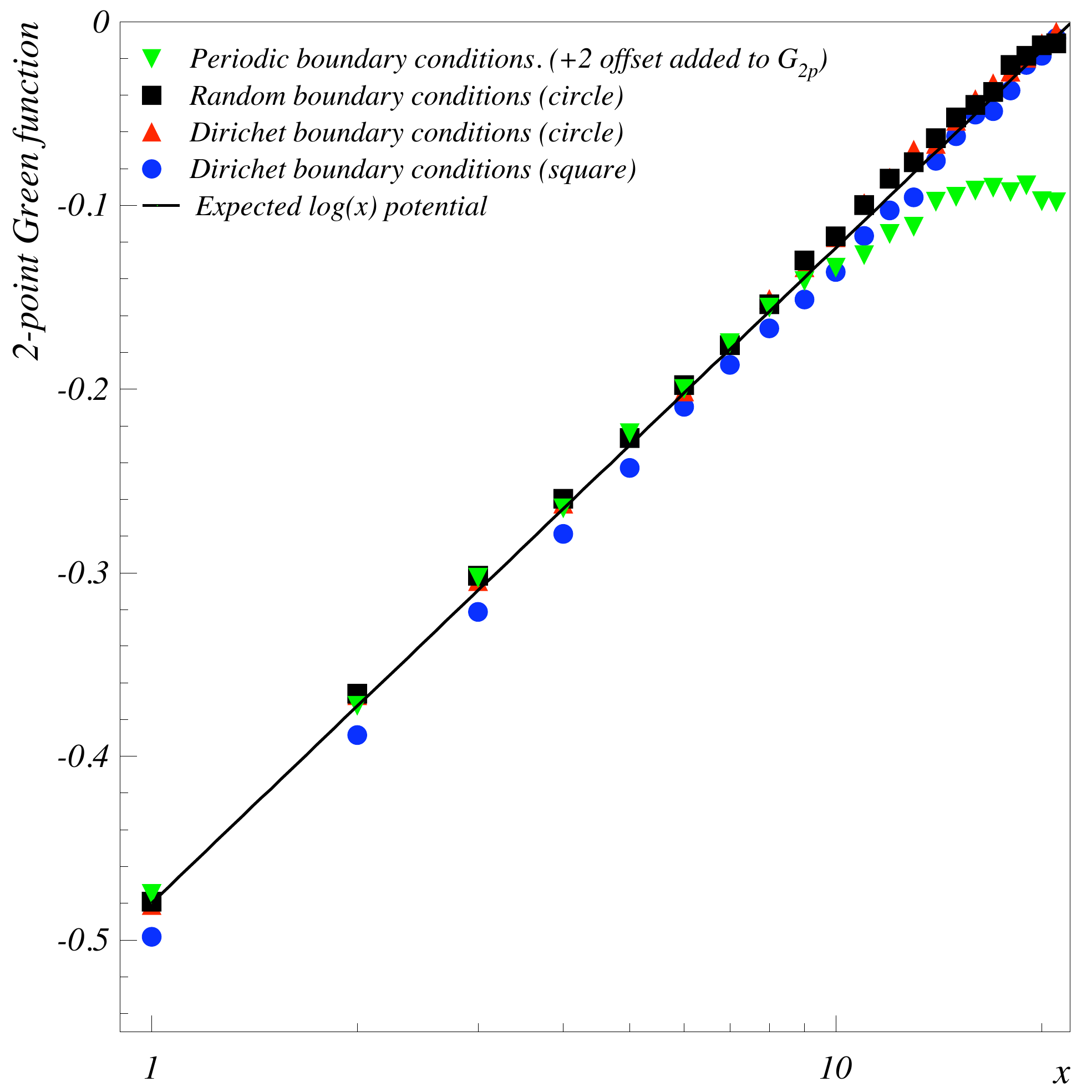}
\caption{
\label{fig: 2D abelian case}
 Potentials in a two-dimensional space obtained when
keeping only the $\partial\phi\partial\phi$ term in the Lagrangian.
In this semi-log graph, the expected logarithmic potential is rectilinear.
The \emph{various symbols} represent different boundary conditions; see main
text \protect \\
}
\end{figure}

We verify here that in two-dimensional space, a logarithmic potential
is recovered in the case of $g=0$ (now integrating the Lagrangian
density only over two dimensions). The simulation was
performed with $N=45$, $N_{\mathrm{cor}}=15$ and $N_{\mathrm{s}}=2.5\times10^{4}$.
We checked the dependence on field values at the boundaries by performing
the calculation for four different cases: 
\begin{enumerate}
\item Dirichlet boundary conditions on the edges of the square lattice.
\item Dirichlet boundary conditions outside a circle of diameter $N$ centered
on the lattice center at $x_{\mathrm{center}}=y_{\mathrm{center}}=N/2$.
\item Random field boundary conditions outside a circle of diameter $N$ centered
on the lattice center.
\item Periodic boundary conditions on the square lattice.
\end{enumerate}
As expected, we obtain $G_{2p}(x)\propto\log (x)$; see Fig.~\ref{fig: 2D abelian case}.
The agreement breaks down at a distance of about 10 lattice units
from the boundary in the case of periodic boundary conditions. This
is because for indices $(i,j)$, $\phi(x=i,y=j)$ is the same as
$\phi(x=i+N,y=j+N)$, i.e. they are fully correlated. Consequently,
$G_{2p}$ is periodic with a period $N$. This effect is not seen
in three dimensions where the correlation quickly falls as $1/x$.
The effect is important for two dimensions where the correlation falls
more slowly, as $\log (x)$. This underlines the importance of choosing
the boundary conditions with care, so that the simulation result is
as independent as possible of the field values at the boundaries.
Clearly, simple periodic boundary conditions (i.e. without
the buffer region discussed in Appendix~\ref{Appendix boundary conditions})
are ill-suited in the case of strongly autocorrelated fields, such
as in the present two-dimensional case and, a fortiori, the
three-dimensional case with large field self-interaction terms.

\section{\label{Appendix lattice param}Dependence on lattice parameters}
The lattice parameters $N$ (lattice size), $N_{\mathrm{s}}$
(statistics), $N_{\mathrm{cor}}$ (decorrelation parameter; see 
Appendix~\ref{sub:First-simulation}) and the lattice spacing $a$ (distance
between adjacent nodes) do not correspond to any physical property
of a system. Results should consequently be independent of them.
We verify this in this appendix,  first for the quadratic case (massless and massive free-fields) and then for the $g^2 \phi^4$ 
and ($g \phi \partial_\mu \phi \partial^\mu \phi + g^2 \phi^2 \partial_\mu \phi \partial^\mu \phi$)  cases.

\subsection{Dependence on the system size \label{sub:Dependence with system size}}
\begin{figure}[ht!] 
\centering
 \includegraphics[width=0.5\textwidth]{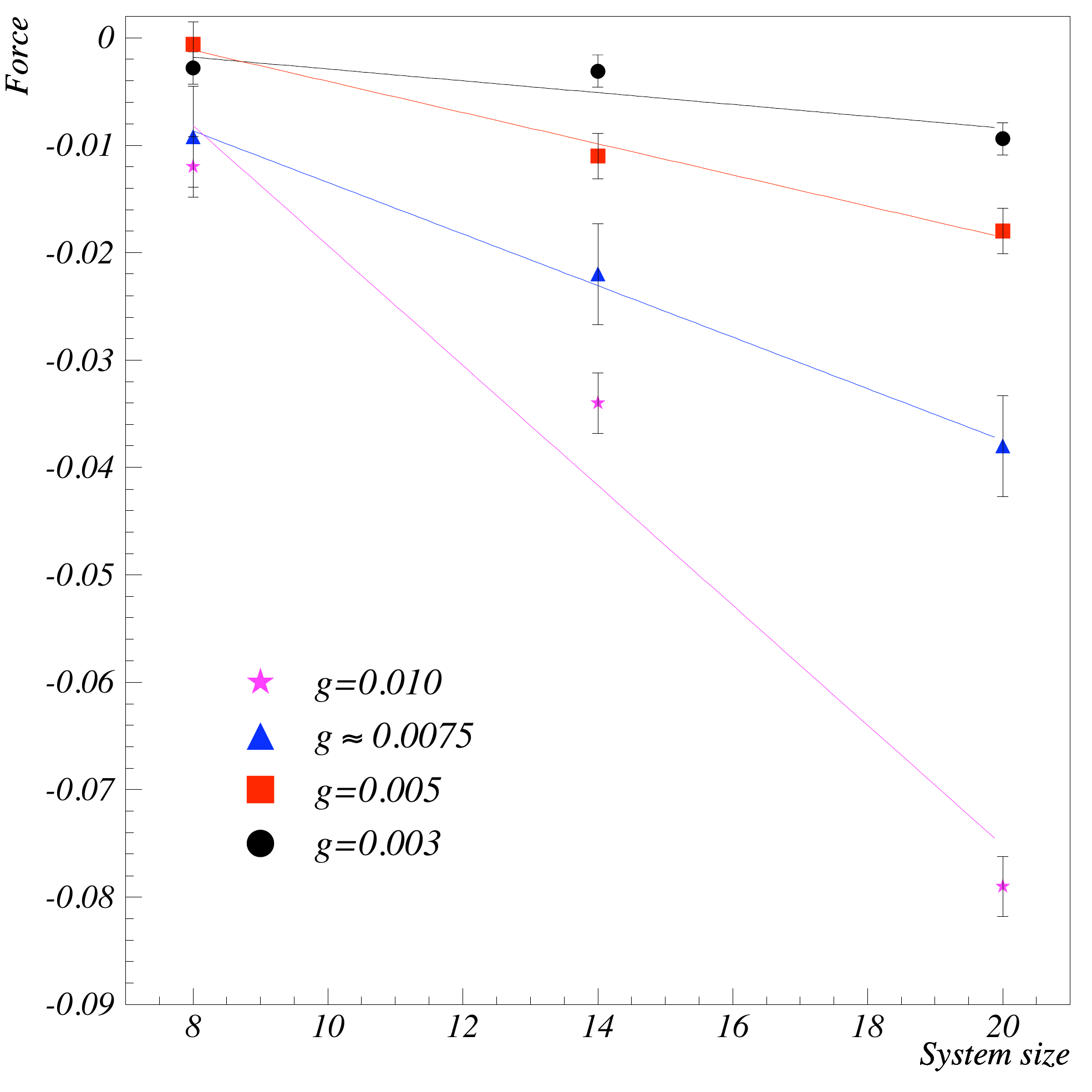}
\caption{
\label{Flo: system size dependence} Dependence of the force between
two sources with the distance $L=2d$ separating them (system size). The \emph{four
different symbols} are results for different values of the 
coupling $g$. (Details of these
calculations are discussed in Sect.~\ref{sub:GR-case}.)
Points are fit with a linear function, yielding an overall $\chi^2/ndf$ of 1.4. The uncertainties
are taken as the difference between the force extracted near source
1 and the force near source 2 \protect \\
}
\end{figure}

A long correlation length implies long-range effects. If the correlation length is 
similar to the system size, these long range effects should depend
on the physical size of the system, or equivalently on the distance
at which the field becomes negligible, since these two scales must
be related. See Ref.~\cite{Hoyer:2009ep} for an example of such phenomenon in an analytical calculation context.
Our calculations for the ($g \phi \partial_\mu \phi \partial^\mu \phi + g^2 \phi^2 \partial_\mu \phi \partial^\mu \phi$) 
theory, Eq.~(\ref{eq:Full Action GR}), do show that the larger the system, the
stronger the long-range effects. This can be seen in Fig.~\ref{Flo: system size dependence}
where the force between two sources separated by a distance $2d$ is
plotted versus $2d$ for the same lattice unit
$a$. The dependence of the force on $2d$ is linear, as expected.

\subsubsection{Free massless and massive field cases\label{sub:verification: Quadratic-cases}}
\begin{figure}[ht!] 
\centering
\includegraphics[scale=0.43]{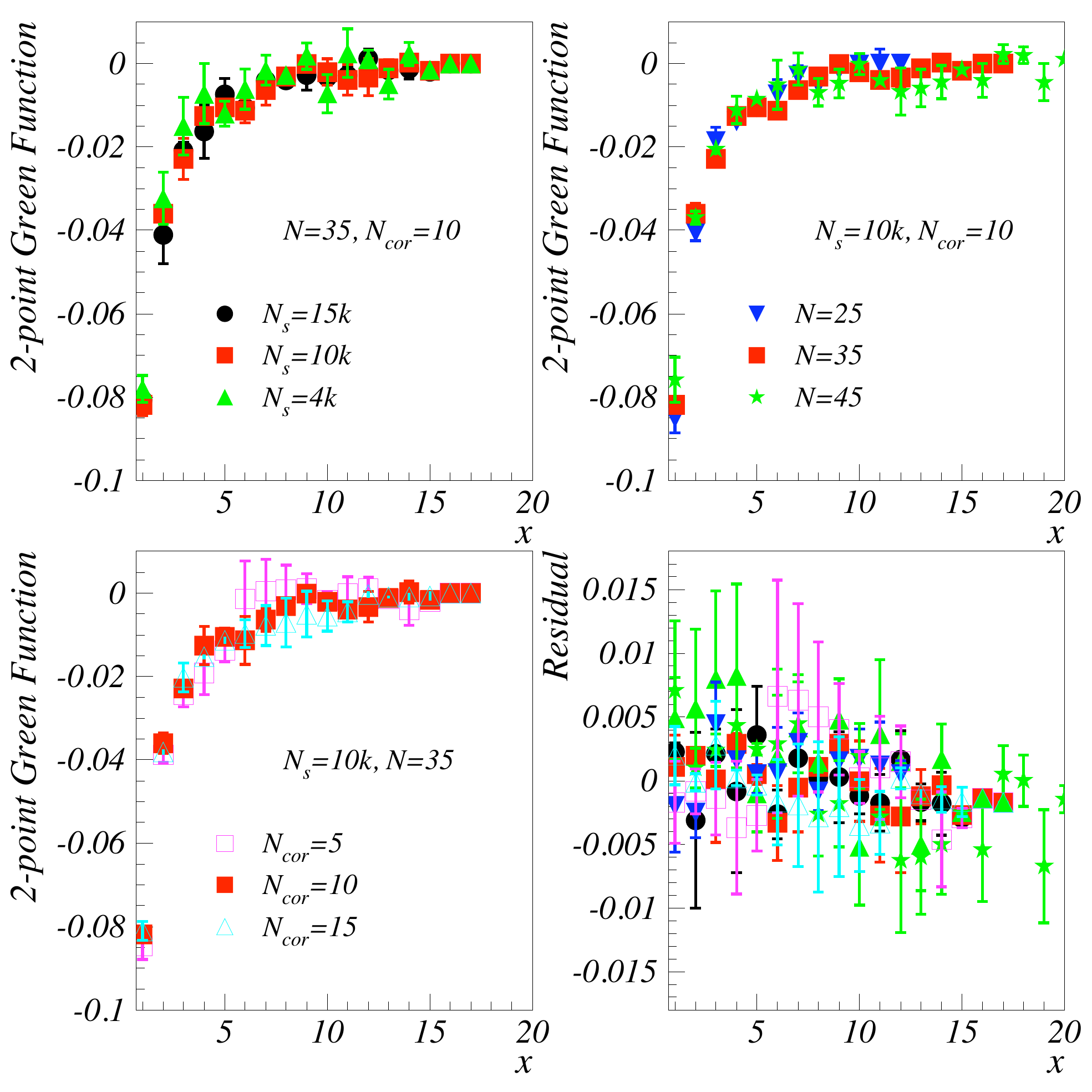}
\caption{\label{fig:param check} Potentials for various values of parameters used
in the simulation. Only the quadratic term $\partial\phi\partial\phi$
in the Lagrangian (massless free-field case) was kept. For these
simulations the system size $N$, the decorrelation parameter $N_{\mathrm{cor}}$,
and the statistics $N_{\mathrm{s}}$, are varied. In the \emph{bottom right
panel}, the residuals between the results and the expected 1/$x$ behavior
are shown. The vertical axis span is the same as for the \emph{three first
panels} and magnified by a factor 3 for the \emph{bottom right panel} showing the residuals \protect \\
}
\end{figure}

We verified in the free massless and massive field cases the independence of $G_{2p}$ on $N$, $N_{\mathrm{cor}}$
and $N_{\mathrm{s}}$; see Fig.~\ref{fig:param check}. In the top
left plot, the effect of statistics ($N_{\mathrm{s}}$) is checked.
In the top right plot, the effect of the lattice size ($N$) is checked.
In the bottom left plot, possible correlations between configurations
($N_{\mathrm{cor}}$) are checked. In the bottom right plot, the residuals
between these results and the expected 1/$x$ behavior are shown.
It appears that $N_{\mathrm{cor}}=10$ is adequate for a simulation
in the quadratic case.

\begin{figure}[ht!]
\centering
\includegraphics[scale=0.37]{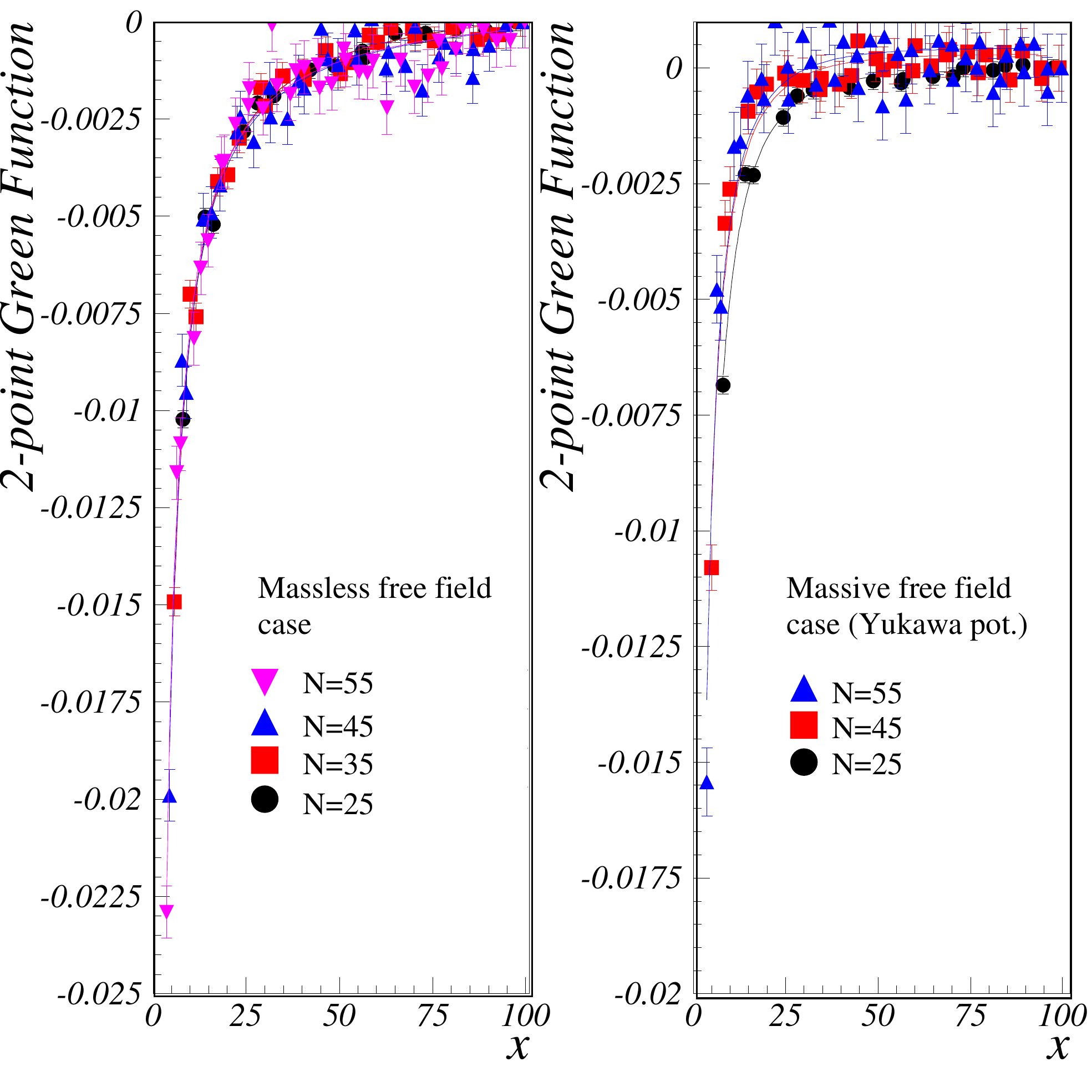}
\caption{\label{fig:scale-abelian}  (Color online) Verification
that $G_{2p}$ calculated numerically is invariant with respect to
changing the lattice spacing units. In the \emph{left panel}, the massless free-field
case ($\mathcal{L}=\partial\phi\partial\phi$) is shown. The case for $\mathcal{L}=\partial\phi\partial\phi+m^{2}\phi^{2}$
(Yukawa potential) is shown on the \emph{right panel}.
Statistics are $N_{\mathrm{s}}=5000$ for results with $N=25,$ $N_{\mathrm{s}}=$4000
for massless free-field results with $N=35$ and $N_{\mathrm{s}}=$2000 for the
other results. A decorrelation parameter of $N_{\mathrm{cor}}$~=~10
has been used. A  mass $m=0.05$ has been used for the Yukawa
potential. The \emph{lines} are the best fits to the data points using the
expected functional forms $1/x$ and $\mathrm{e}^{-mx}/x$ for the
massless and massive  cases, respectively. Their \emph{color} matches the \emph{symbol
colors} }
\end{figure}

The results must also be independent of the physical value of the
lattice spacing unit $a$, i.e. calculations on lattices of different
sizes $N$ but corresponding to the same physical size must yield
the same result. The values of $G_{2p}$ obtained with the simulation
performed for different values of $N$ are plotted on top of each
other in Fig.~\ref{fig:scale-abelian}. Their
respective $x-$dependencies are scaled by a factor $L/N$, where
$L$ is the system physical size, taken to be 100 in Fig.~\ref{fig:scale-abelian}.
The dimension of {[}$G_{2p}(n)${]} is $[\phi]^{2}=[\mbox{length}]^{-1}$,
so $G_{2p}$ must also be scaled by a factor $N/L$.
Finally, if a  mass term is present (Yukawa case), it must
be scaled in the simulation by a factor $N/L$, since $m$ scales
as {[}length{]}$^{-1}$. The left panel in Fig.~\ref{fig:scale-abelian}
shows the massless free-field case and the right panel the Yukawa case. In both
instances, the calculations for different $N$ values agree well. We next verify the independence 
from the lattice parameters, first in the  $g^2 \phi^4$ theory and then in the  case of the
($g \phi \partial_\mu \phi \partial^\mu \phi + g^2 \phi^2 \partial_\mu \phi \partial^\mu \phi$)  theory. 

\subsubsection{$g^2 \phi^4$ case\label{sub:QCD-case}}
\begin{figure}[ht!] 
\centering
\includegraphics[scale=0.43]{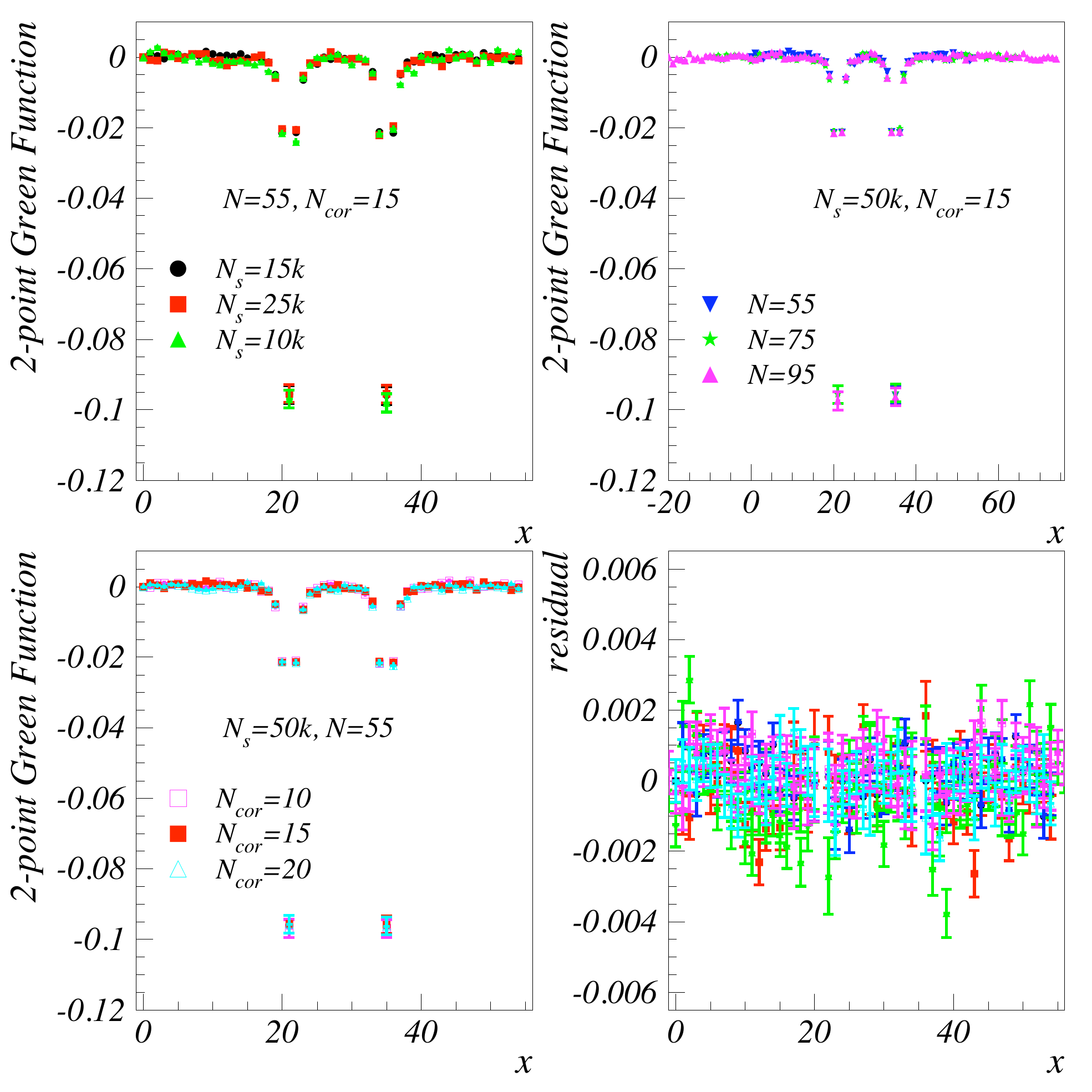}
\caption{\label{fig:param check QCD} Potentials for various values of parameters used
in the simulation. The $g^2 \phi^4$ Lagrangian, Eq.~(\ref{eq:Toy Lagrangian QCD}), is used with $g^2=0.5$. 
The system size, $N$, the decorrelation parameter, $N_{\mathrm{cor}}$,
and the statistics, $N_{\mathrm{s}}$, are varied. In the \emph{bottom right
panel}, the residuals between the results and the effective Yukawa behavior
are shown. The \emph{vertical axis} span is the same as for the \emph{three first
panels} and magnified by a factor 10 for the \emph{bottom right panel} showing the residuals\protect \\
}
\end{figure}

The dependence of the results with variation of the lattice size $N$, decorrelation parameter $N_{\mathrm{cor}}$,
and statistics $N_{\mathrm{s}}$ is shown in Fig.~\ref{fig:param check QCD}. We used $g^2=0.5$, which yields an effective 
Yukawa behavior characterized by an effective field mass $m_{\mbox{\scriptsize{eff}}} \simeq 0.72$. There is an excellent agreement 
between the data  indicating no dependence on the choice of lattice parameters.

\subsubsection{($g \phi \partial_\mu \phi \partial^\mu \phi + g^2 \phi^2 \partial_\mu \phi \partial^\mu \phi$)  theory case\label{sub:GR-case}}
\begin{figure}[ht!] 
\centering
 \includegraphics[width=0.5\textwidth]{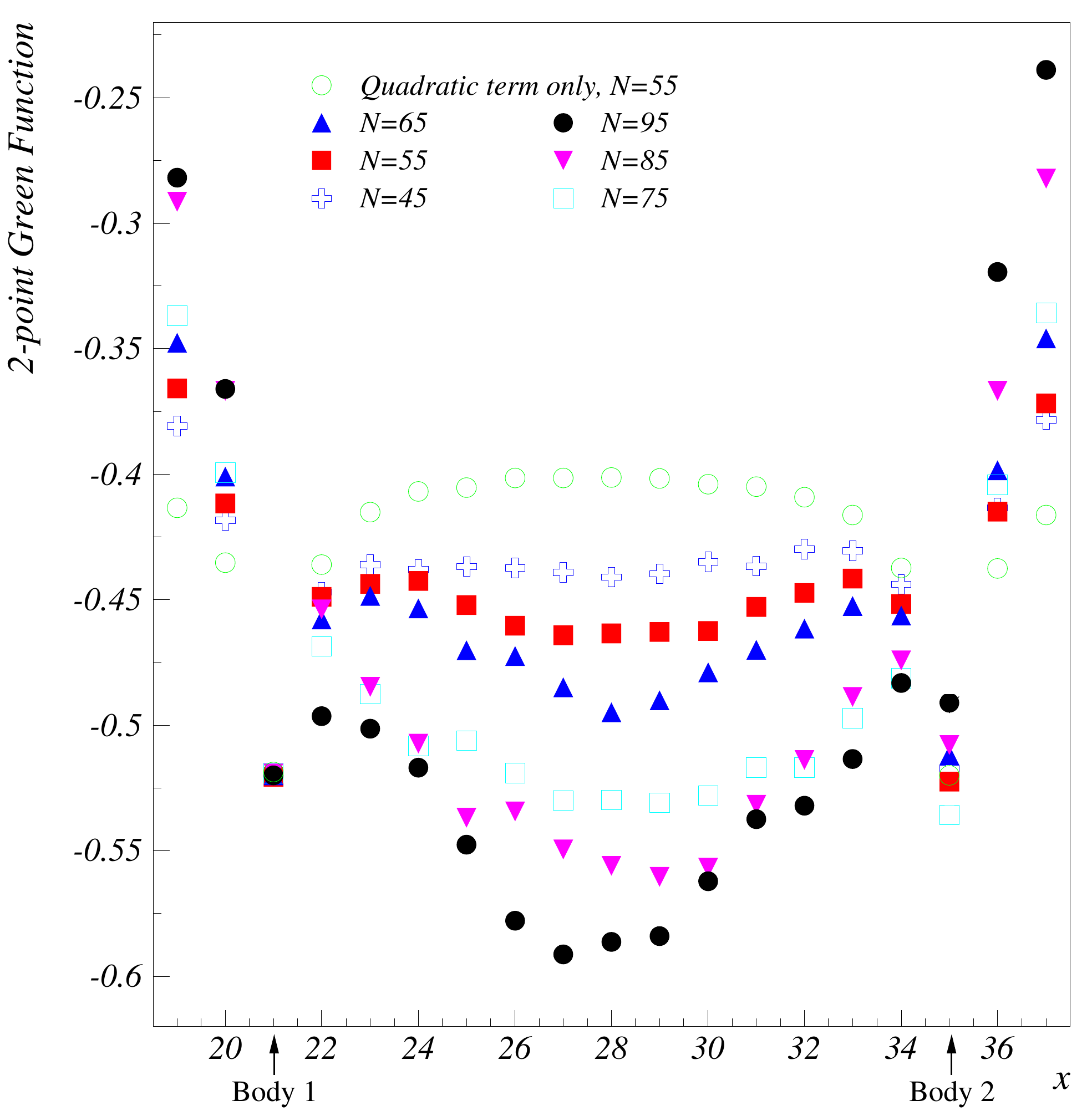}
\caption{\label{Flo:size effect all}(Color online) Potentials computed for different lattice 
sizes varying from $N=45$ to $N=95$. The Lagrangian in Eq.~(\ref{eq:Full Action GR})
is used except for the \emph{open circles} that show the massless free-field case. The two bodies are located on the $x-$axis
at $d=\pm7$ from the lattice center $x_{\mathrm{center}}=28$, $y=0$
and $z=0$. The coupling is $g=7.5\times10^{-3}$. The decorrelation parameter is $N_{\mathrm{cor}}=20$ and 
$N_{\mathrm{s}}=3.5\times10^{4}$ paths were used.
To facilitate comparing the results, a constant offset is added to
the data for each $N$ cases so that the data at the position of the first body, $x=21$, match\protect \\
}
\end{figure}

\begin{figure}[ht!] 
\centering
 \includegraphics[width=0.5\textwidth]{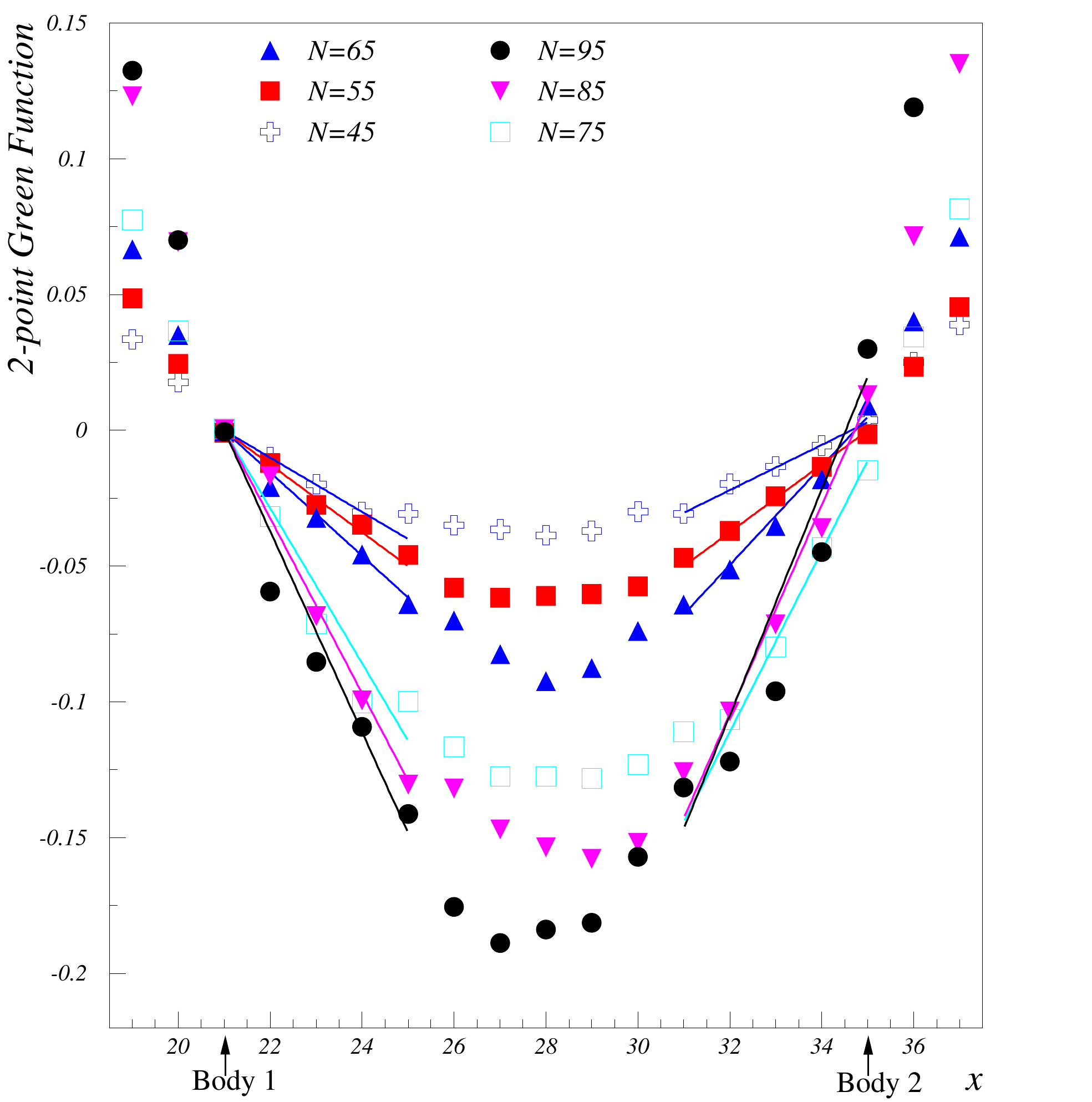}
\caption{\label{Flo:size effect all Newton subtracted}Same as Fig.~\ref{Flo:size effect all}
but with the free-field potential subtracted. The line segments
illustrate the nearly linear trend of the potential\protect \\
}
\end{figure}

\paragraph{Dependence on lattice size.}
In Fig.~\ref{Flo:size effect all}, we show $G_{2p}(x)$ computed for different
lattice sizes varying from $N=45$ to $N=95$. Dirichlet boundary
conditions were used. The distance between the two sources is $2d=14$ lattice
spacings for all $N$. For ease of comparison, arbitrary constants
were added to the $G_{2p}(x)$s so that the troughs at distance $x=21$
in the top panel overlap. (Constants do not affect the forces resulting
from the potentials.) The effect of the cubic and quartic terms on the force is obtained by taking
the derivative of $G_{2p}$ after subtracting the free-field
contribution (open circles). There is a $N-$dependence of the force.
It is approximately linear with $N$ near the matter locations within the $N-$range checked
($45 \leq N \leq 115$) for the various cases checked ($d=4,$ 7 and 10, and
$3\times10^{-3}\leq g \leq 10^{-2}$); see Fig.~\ref{Flo:size effect all Newton subtracted}.
This difference in slope between the results for different  $N$ is 
likely a residual effect similar to the physical effect
discussed in Sect.~\ref{sub:Dependence with system size} but 
associated with the buffer region size or the lattice size rather than the physical size. 
Consequently, the induced linear $N-$dependence
must be corrected for. It can be done, for example, by subtracting
the $N-$dependent component. Practically, it is done by taking the
$N\rightarrow0$ (classical) limit.

\begin{figure}[ht!] 
\centering
 \includegraphics[width=0.6\textwidth]{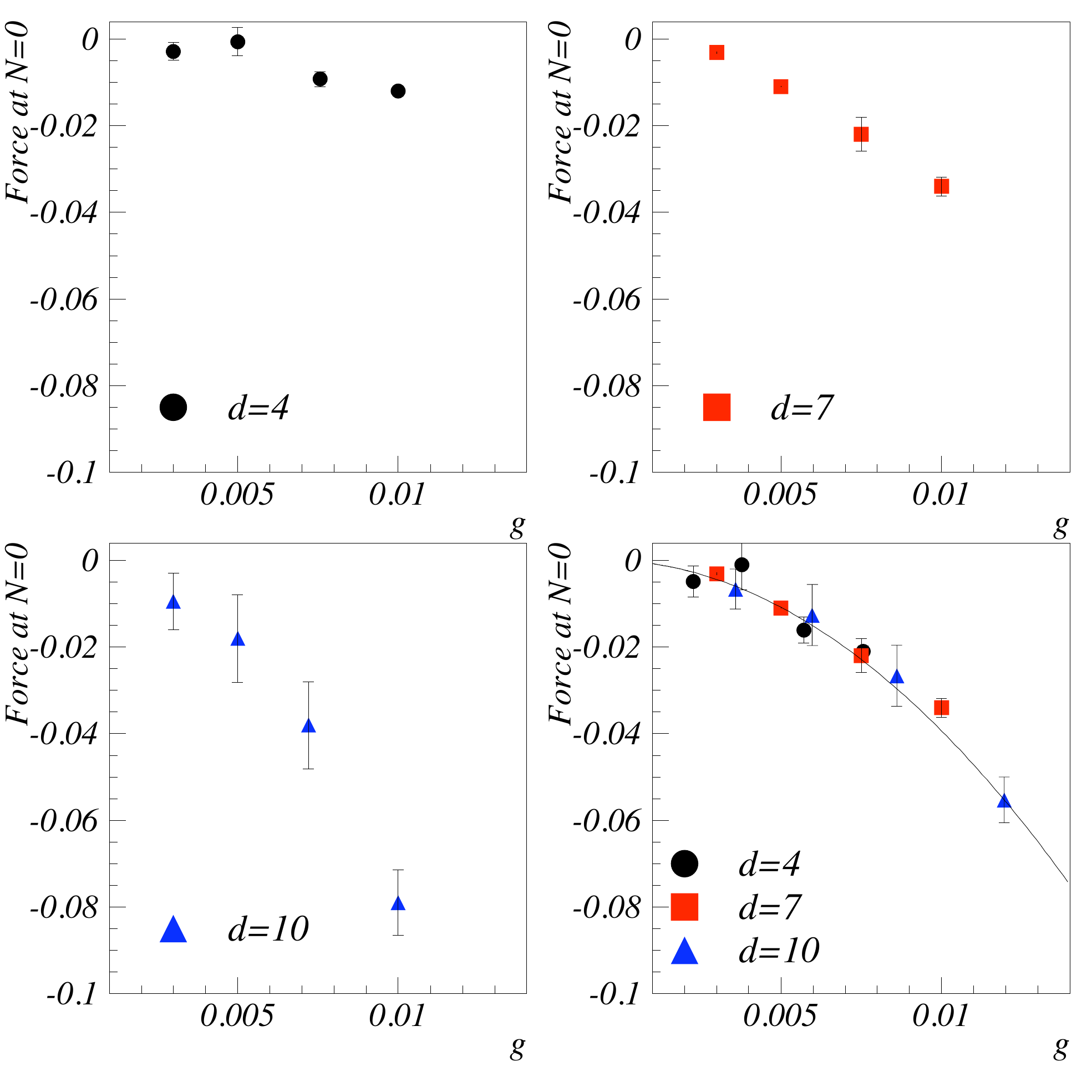}
\caption{
\label{Flo:lattice spacing check} Dependence of the
simulation on lattice spacing. The Lagrangian of Eq.~(\ref{eq:Full Action GR})
is used. In the \emph{four panels}, the $\mathrm{d}(G_{2p}(x)-G_{2p}^{\mathrm{free-field}}(x))/\mathrm{d}x$ part of the force along
$x$ is plotted as a function of the coupling $g$, in the classical $N\rightarrow0$
limit. The number of lattice nodes between the two matter terms is
$2d=8$ (\emph{top left panel, circles}), $2d=14$ (\emph{top right panel,
squares}) and $2d=20$ (\emph{bottom left panel, triangles}). The
\emph{bottom right plot} shows these results for a same
physical two-source system, after properly rescaling for the difference
in physical lattice units. The three calculations using different
physical lattice units agree well. The \emph{full curve} is a second-order
polynomial fit to the data\protect \\
}
\end{figure}

\paragraph{Dependence on lattice parameters}
As shown in Appendix~\ref{Appendix boundary conditions}, the results are
mostly independent of the choice of field values at the boundaries.
We verify now that the results are also independent of the lattice
spacing if $g$ is small enough and the boundaries are sufficiently
far from the points where $G_{2p}$ is calculated. The effect of the cubic term $g \phi \partial_\mu \phi \partial^\mu \phi$  and 
quartic term $g^2 \phi^2 \partial_\mu \phi \partial^\mu \phi$ on the force between the two sources, in the classical $N\rightarrow0$ limit,
is shown in Fig.~\ref{Flo:lattice spacing check}. It is given along
$x$ by $\mathrm{d}G_{2p}(x)/\mathrm{d}x$ and is measured a few nodes
away from the sources (e.g. around $x=21$ and $x=35$ in Fig.~\ref{Flo:size effect all Newton subtracted}).
The free massless field  contribution $G_{2p}^{\mathrm{free-field}}(x)\propto[1/(x-x_{1})+1/(x-x_{2})]$
was first subtracted. (We used the lattice result obtained with $g^2=0$
for the subtraction rather than the analytic expression.) The simulation
was performed for lattice sizes $45 \leq N \leq 95$, for three different numbers
of lattice nodes between the two sources: $2d=8,$ 14 and 20, and for
couplings $3\times10^{-3}<g<10^{-2}$. The correlation parameter was
$N_{\mathrm{cor}}=20$, the statistic $N_{\mathrm{s}}=3.5\times10^{4}$,
and the Dirichlet boundary conditions
were used. For the same two-source system, the physical lattice unit
$a$ is a factor 7/4 times larger for the $d=4$ calculation compared
to the $d=7$ calculation, and a factor 7/10 times smaller for the
$d=10$ calculation compared to $d=7$. Since the dimension of $g$
is $\sqrt{[\mbox{length]}}$ and the force dimension is $[\mbox{length]}^{-2}$,
the respective couplings must be scaled as $g{}_{d_{1}}=g{}_{d_{2}}\sqrt{d_{1}/d_{2}}$
in order to compare a calculation with $d=d_{1}$ to one with $d=d_{2}$.
The respective forces must be scaled on the one hand by $(d_{2}/d_{1})^{2}$ due to their distance$^{-2}$ dimension,
but on the other hand, divided by $(d_{2}/d_{1})$ to account for
the intrinsic linear $N-$dependence discussed in Sect.~\ref{sub:GR-case},
so that overall $F_{d_{1}}=F{}_{d_{2}}d_{2}/d_{1}$. Once the couplings
and forces are thus scaled to account for their non-zero dimensions, the calculations performed using different
values of $a$ agree well; see the bottom right panel of Fig.~\ref{Flo:lattice spacing check}.
(This invariance with the lattice unit is valid independent of taking the $N\rightarrow0$ limit.)

\section{\label{Appendix: Spin 0 Assumption} Recovering the post-Newtonian
formalism}
We discussed in Sect.~\ref{sec: Static Lagrangian} how scalars
fields could describe static systems. To verify that Eq.~(\ref{eq:Full Action GR}) provides an adequate
static approximation of GR, we retrieve here the weak-field perturbative
post-Newtonian formalism from the scalar Lagrangian in the static
case of two bodies. We start from the Lagrangian of
Eq.~(\ref{eq:Full Action GR}) truncated to first
order in $\sqrt{16\pi G}$:
\begin{eqnarray}
\mathcal{L}={\bigl(\nabla\phi\bigr)^{2}+} \sqrt{16\pi G}\phi\bigl(\nabla\phi\bigr)^{2} \label{eq:ppn}
-\sqrt{16\pi G}\phi\bigl(M_{1}\delta^{(3)}(r-r_{1})+M_{2}\delta^{(3)}(r-r_{2})\bigr),\end{eqnarray}
\noindent where $r$ is the position of a test mass and $r_{1}$
and $r_{2}$ the positions of the two bodies of masses $M_{1}$ and
$M_{2}$, respectively. The Euler-Lagrange equations yield
\begin{equation}
\Delta\phi=-\frac{\sqrt{16\pi G}}{2}\frac{M_{1}\delta^{(3)}(r-r_{1})+M_{2}\delta^{(3)}(r-r_{2})+\left(\nabla\phi\right)^{2}}{1+\sqrt{16\pi G}\phi}.\label{eq:Poisson eq.}\end{equation}
\noindent Assuming a weak field, $\sqrt{16\pi G}\phi\ll1$, and that
the relative change of $\phi$ are small at long distances, i.e.
$\nabla\phi\lesssim\phi$ (this is true for the zero order Newton
solution and is justified \emph{a posteriori} for the first order
post-Newtonian solution), we can solve Eq.~(\ref{eq:Poisson eq.})
to first order in $\sqrt{16\pi G}\phi$:
\begin{eqnarray}
\phi=-\frac{\sqrt{16\pi G}M_{1}}{2(r-r_{1})} -\frac{\sqrt{16\pi G}M_{2}}{2(r-r_{2})}+\frac{(16\pi G)^{3/2}}{4} 
\left(\frac{M_{1}^{2}}{r-r_{1}}+\frac{M_{2}^{2}}{r-r_{2}}+2\frac{M_{1}M_{2}}{(r-r_{1})(r-r_{2})}\right).\end{eqnarray}
\noindent In the center of mass frame and suppressing the self-energy
terms, we obtain the post-Newtonian potential energy in the static
case:
\begin{equation}
E=\frac{GM_{1}M_{2}}{r}\left(1-\frac{G(M_{1}+M_{2})}{2r}\right). \label{eq:pnp}
\end{equation}
\noindent (To obtain Eq.~\ref{eq:pnp}, one must use the standard
field normalization, $\phi\equiv(g_{\mu\nu}-\eta_{\mu\nu})$, i.e.
replacing $\phi$ by $\phi/\sqrt{16\pi G}$.) Many
strong field effects are non-perturbative and cannot be obtained with
such perturbative formalism. However, the exercise above validates
the use of Eq.~(\ref{eq:Full Action GR}) for static
non-perturbative calculations.

\end{document}